\documentclass[aps,prd,a4paper,twocolumn,amsmath,showpacs,superscriptaddress,nofootinbib,preprintnumbers]{revtex4-1}

\usepackage{graphicx,ulem}
\usepackage{longtable}
\usepackage{float}
\usepackage{dcolumn}
\usepackage{graphics,epsfig}
\usepackage{soul}
\usepackage{amsmath,amssymb,latexsym,mathrsfs}
\usepackage{bm}
\usepackage{color}
\usepackage{color}
\usepackage{subfigure}
\usepackage{multirow}

\newcommand{\Pscalar}{\ensuremath{P_\mathcal{R}}}
\newcommand{\Pscalarrt}{\ensuremath{\Pscalar^{1/2}}}
\newcommand{\Ptensor}{\ensuremath{P_T}}
\newcommand{\Ptensorrt}{\ensuremath{\Ptensor^{1/2}}}

\newcommand{\mpl}{\ensuremath{m_\textrm{Pl}}}
\newcommand{\mgut}{\ensuremath{m_\textrm{GUT}}}

\newcommand{\ns}{\ensuremath{n_s}}
\newcommand{\nt}{\ensuremath{n_\textrm{T}}}

\newcommand{\eV}{\,\mathrm{eV}}

\newcommand{\Neff}{N_\mathrm{eff}}

\newcommand{\lcdm}{\Lambda\mathrm{CDM}}

\def\he4{$^4$He}
\def\h2{$^2$H}

\def\aap{\ref@jnl{A\&A}}                

\begin{document}
\preprint{NORDITA-2016-113,IFIC/16-75}

\title{Impact of neutrino properties on the estimation of inflationary parameters from current and future observations}

\author{Martina Gerbino}
\email{martina.gerbino@fysik.su.se}
\affiliation{The Oskar Klein Centre for Cosmoparticle Physics, Department of Physics, Stockholm University, AlbaNova, SE-106 91 Stockholm, Sweden}
\affiliation{The Nordic Institute for Theoretical Physics (NORDITA), Roslagstullsbacken 23, SE-106 91 Stockholm, Sweden}

\author{Katherine Freese}
\email{ktfreese@umich.edu}
\affiliation{The Oskar Klein Centre for Cosmoparticle Physics, Department of Physics, Stockholm University, AlbaNova, SE-106 91 Stockholm, Sweden}
\affiliation{The Nordic Institute for Theoretical Physics (NORDITA), Roslagstullsbacken 23, SE-106 91 Stockholm, Sweden}
\affiliation{Michigan Center for Theoretical Physics, Department of Physics, University of Michigan, Ann Arbor, MI 48109, USA}

\author{Sunny Vagnozzi}
\email{sunny.vagnozzi@fysik.su.se}
\affiliation{The Oskar Klein Centre for Cosmoparticle Physics, Department of Physics, Stockholm University, AlbaNova, SE-106 91 Stockholm, Sweden}
\affiliation{The Nordic Institute for Theoretical Physics (NORDITA), Roslagstullsbacken 23, SE-106 91 Stockholm, Sweden}

\author{Massimiliano Lattanzi}
\affiliation{Dipartimento di Fisica e Scienze della Terra, Universit\`a di Ferrara and INFN, Sezione di Ferrara, Polo Scientifico e Tecnologico - Edificio C Via Saragat, 1, I-44122 Ferrara, Italy}

\author{Olga Mena}
\affiliation{IFIC, Universidad de Valencia-CSIC, 46071, Valencia, Spain}

\author{Elena Giusarma}
\affiliation{McWilliams Center for Cosmology, Department of Physics, Carnegie Mellon University, Pittsburgh, PA 15213, USA}
\affiliation{Lawrence Berkeley National Laboratory (LBNL), Physics Division, Berkeley, CA 94720-8153, USA}

\author{Shirley Ho}
\affiliation{McWilliams Center for Cosmology, Department of Physics, Carnegie Mellon University, Pittsburgh, PA 15213, USA}
\affiliation{Lawrence Berkeley National Laboratory (LBNL), Physics Division, Berkeley, CA 94720-8153, USA}
\affiliation{Departments of Physics and Astronomy, University of California, Berkeley, CA 94720, USA}

\date{\today}

\begin{abstract}
We study the impact of assumptions about neutrino properties on the estimation of inflationary parameters from cosmological data, with a specific focus on the allowed contours in the $n_s/r$ plane, where $n_s$ is the scalar spectral index and $r$ is the tensor to scalar ratio. 
We study the following neutrino properties:
(i) the total neutrino mass $ M_\nu =\sum_i m_i$ (where the index $i = 1,\,2,\,3$ runs over the three neutrino mass eigenstates); (ii) the number of relativistic degrees of freedom $\Neff$ at the time of recombination; and (iii) the neutrino hierarchy: whereas previous literature assumed 3 degenerate neutrino masses or two massless neutrino species (approximations that clearly do not match neutrino oscillation data), we study the cases of normal and inverted hierarchy.  Our basic result is that these three neutrino properties induce $< 1 \sigma$ shift of the probability contours in the $\ns/r$ plane with both current or
upcoming data.
We find that the choice of neutrino hierarchy (normal, inverted, or degenerate)
has a negligible impact. 
  However, the minimal cutoff on the total neutrino mass $M_{\nu,{min}}=0 $ that accompanies previous works using the degenerate hierarchy does introduce biases in the $\ns/r$ plane and should be replaced by $M_{\nu,\mathrm{min}}= 0.059\,\eV$ as required by oscillation data.
Using current Cosmic Microwave Background (CMB) data from Planck and Bicep/Keck (BK14),  marginalizing over the total neutrino mass $ M_\nu$ and over $r$ can lead to a shift in the mean value of $n_s$ of $\sim0.3\sigma$ towards lower values. However, once Baryon Acoustic Oscillation (BAO) measurements are included, the standard contours in the $n_s/r$ plane are basically reproduced.  Larger shifts of the contours in the $n_s/r$ plane (up to 0.8$\sigma$) arise for nonstandard values of $\Neff$. We also provide forecasts for the future CMB experiments COrE (satellite) and Stage-IV (ground-based) and show that the incomplete knowledge of neutrino properties, taken into account by a marginalization over $M_\nu$, could induce a shift of $\sim0.4\sigma$ towards lower values in the determination of $n_s$  (or a $\sim 0.8\sigma$ shift if one marginalizes over $\Neff$).     Comparison to specific inflationary models is shown.  Imperfect knowledge of neutrino properties
must be taken into account properly, given the desired precision in determining whether or not inflationary models match the future data.

\end{abstract}

\maketitle

\section{Introduction}\label{sec:intro}
Inflation, which consists of a period of accelerated expansion prior to the conventional radiation and matter dominated epochs, provides the most compelling framework to address the homogeneity, flatness, and monopole problems  of the standard Big Bang cosmology (see e.g. \cite{Guth:1980zm,Starobinsky:1980te,Mukhanov:1981xt,Linde:1981mu,Albrecht:1982wi} for pioneering work, and \cite{Kazanas:1980tx,Sato:1981ds} for important earlier work).  
 Over the past decade Cosmic Microwave Background (CMB) observations
have confirmed basic predictions of inflation and in addition have provided stringent tests 
of individual inflationary models.
  First, generic predictions of inflation
match the observations: the universe has a critical density
($\Omega=1$), the density perturbation spectrum is nearly scale
invariant, and superhorizon fluctuations are evident. Second, current
data  differentiate between inflationary models and
rule some of them out \cite{Spergel:2006hy,Alabidi:2006qa,Peiris:2006ug,Peiris:2006sj,Easther:2006tv,Seljak:2006bg,
Kinney:2006qm,Martin:2006rs,Martin:2013tda,Martin:2013nzq,Planck:2013jfk,Ade:2015lrj,Huang:2015cke}.
 
 Inflation models predict two types of perturbations, scalar and tensor,
which result in density and gravitational wave fluctuations,
respectively.  Each of them is typically characterized by a fluctuation
amplitude ($\Pscalarrt$ for scalar and $\Ptensorrt$ for tensor, with
the latter usually given in terms of the ratio $r \equiv
\Ptensor/\Pscalar$) and a spectral index ($\ns$ for scalar and
$\nt$ for tensor) describing the mild scale dependence of the
fluctuation amplitudes.  The amplitude $\Pscalarrt$ is normalized
by the height of the inflationary potential. For single-field slow-roll models, the inflationary
consistency condition $r = -8 \nt$ further reduces the number of free
parameters to two, leaving experimental limits on $\ns$ and $r$ as the
primary means of distinguishing among inflationary models.  Hence,
predictions of models are presented as plots in the $\ns/r$ plane.

Previous works have investigated the shifts induced in the $\ns$ and/or $r$ inflationary parameters caused by our poor understanding of the early-universe physics (\textit{i.e.} the reheating process~\cite{Munoz:2014eqa,Dai:2014jja}) or the late-time universe evolution (as, for instance, the precise reionization details~\cite{Mortonson:2007tb,Pandolfi:2010dz,Pandolfi:2010mv,Dizgah:2012em,Oldengott:2016yjc}). 

In this context, neutrino properties, which play a role in both early and late stages of our universe, show important correlations with the inflationary parameters \cite{Canac:2016smv, Oyama:2015gma,  Ade:2015lrj, Planck:2013jfk, Takada:2005si, Carbone:2010ik}.  Indeed assumptions about the neutrino properties may bias the extraction of inflationary parameters and hence lead to incorrect conclusions about which inflationary potentials match the data \cite{Tram:2016rcw}.
In this paper we study the effects of three neutrino properties that are particularly important in this regard: the neutrino hierarchy, the total neutrino mass, and the contribution of neutrinos to the total radiation content through the effective number of relativistic degrees of freedom at recombination $\Neff$.

{\it Neutrino Hierarchy:}
While from neutrino oscillation experiments we know that at least two out of the three standard model active neutrino species are massive, we do not know their precise hierarchical structure nor the total absolute value of the neutrino masses, as oscillation measurements only provide information on the splittings between the three neutrino mass eigenstates, see e.g. Refs.~\cite{Forero:2014bxa,nufit} and references therein for the most recent global fit analyses.
 The sign of the largest mass splitting, the atmospheric mass gap $\Delta m^2_{\rm{atm}}$, remains unknown. The two possibilities, $\Delta m^2_{\rm{atm}}>0$ and $\Delta m^2_{\rm{atm}}<0$, have been dubbed as Normal (NH) and Inverted (IH) hierarchies, respectively. We will consider both possibilities. We shall compare this novel approach to previous approximations, described in what follows.
 
 Previous work in the literature made the simplifying assumptions of (a) two massless neutrinos (a proxy for the Normal Hierarchy case when the total neutrino mass is fixed to the minimal value allowed by oscillation experiments) or (b) three degenerate neutrinos (which is a good approximation as long as the total neutrino mass is much higher than the minimal mass, see e.g.~\cite{Lesgourgues:2004ps}).  Clearly these assumptions
 do not match the real world, since we know from oscillation experiments that at least two of the 
 active species are massive. Hence it is the goal of this paper to reexamine constraints on inflation, with the measured mass splittings, studying precisely the NH and IH scenarios. 

{\it Total Neutrino Mass:}
\hfill\break
A combination of cosmological measurements including Cosmic Microwave Background (CMB) and Baryon Acoustic Oscillations (BAO) provides currently the tightest constraints on the sum of the masses of the three active massive neutrinos, setting $\sum m_\nu < 0.214\,\eV$ at 95\%~confidence level (CL).~This upper bound results from Planck 2015 full temperature and large scale polarization in combination with BAO measurements, assuming a one-parameter extension of the standard $\lcdm$ model, the addition of the sum of 
neutrino masses \footnote{See \url{http://www.cosmos.esa.int/web/planck/pla} for further details. See e.g.~\cite{Ade:2015xua} or \cite{Aghanim:2016yuo} for the updated constraints based on recent estimates of the reionization optical depth $\tau$ from Planck HFI. The quoted bound can be further tightened down to $\approx$ 0.12-0.15 eV by adding power spectrum (matter or Ly-$\alpha$) data and/or by considering a prior on the Hubble parameter, see e.g. \cite{Palanque-Delabrouille:2014jca,Cuesta:2015iho,Huang:2015wrx,Giusarma:2016phn,Hannestad:2016fog,Bonvin:2016crt,
Alam:2016hwk}.}.  However, these bounds have made a few assumptions, which we relax in this paper. In particular, they  have been obtained by assuming three degenerate neutrinos of equal mass.  This assumption clearly does not match completely the reality as at least two of the neutrinos are known to have mass, yet it may be adequate given
current data; we will test the acceptability of this assumption.  Second, these bounds assume $r=0$.  
In addition, these bounds are obtained for the case of standard $\Lambda$CDM supplemented by one single parameter in 
that the total neutrino mass is marginalized over. In this paper, we will vary a number of additional parameters as well as 
relaxing the assumptions inherent in the above bounds. 

Previous works (see e.g.~\cite{Ade:2015lrj, Planck:2013jfk}) have studied shifts in the 
$\ns/r$ plane due to marginalizing over the sum of neutrino masses and $r$; our contribution
is to redo this study with the correct neutrino hierarchies taken into account, as well as using the latest available data.

\textit{Number of Relativistic Species at Recombination:}
Third, neutrino abundances, settled around the Big Bang Nucleosynthesis (BBN) period, may differ from their canonical expectation. 
If any of the neutrino species contributes a different amount to the radiation content of the Universe, then the epoch of matter/radiation 
equality may shift, leading to shifts in the predictions for $n_s$ that would cause misinterpretation of CMB data in terms of 
underlying inflationary models. The number of relativistic degrees of freedom $\Neff$ at the time of recombination (\textit{i.e.} the quantity which the CMB is sensitive to) may
serve as a proxy for this effect.  The standard
neutrino contribution predicts $\Neff=3.046$~\cite{Mangano:2005cc}.  
Current cosmological data analyses result in a  value of $\Neff=3.15\pm0.23$~\cite{Ade:2015xua} \footnote{This constraint arises from the combination of Planck 2015 full temperature data, large-scale polarisation and BAO measurements in the context of a $\lcdm$ model supplemented by the one parameter extension given by marginalizing over $\Neff$. A full list of constraints from additional combinations of datasets can be found in the Planck Legacy Archive at \url{http://www.cosmos.esa.int/web/planck/pla}.}. However, the predictions may vary in either direction. Somewhat extreme low-reheating scenarios have been proposed~\cite{Kawasaki:1999na,Kawasaki:2000en,Giudice:2000ex,Giudice:2000dp}, where the reheating temperature $T_\mathrm{RH}$ can be as low as 5 MeV \cite{deSalas:2015glj}.  In low-reheating scenarios the neutrino populations will not reach the expected thermal abundances, leading to values of $\Neff<3.046$. Sterile neutrino species, axions, hidden photons, or any other extra dark radiation species could instead lead to a value $\Neff>3.046$  \cite{Archidiacono:2013fha}.  Modified predictions to $\Neff$ shift the predicted $\ns$, implying non-negligible consequences for some inflationary models.  Previous works~\cite{Tram:2016rcw,Ade:2015lrj,Planck:2013jfk} studied the shift due to the 
addition of dark radiation with the assumptions of two massless and one massive neutrino.  We
treat also the case of $\Neff\leq3.046$ in low reheating scenarios.

The assumed fiducial values of the three crucial neutrino properties listed above (namely, the number of relativistic species, the neutrino spectrum, and the total neutrino mass $\sum m_\nu$) could bias the determination of the inflationary parameters and therefore the extraction of the underlying inflationary potential.  It is the main goal of this study to assess the current and future biases from Cosmic Microwave Background (CMB) measurements in the scalar spectral index $n_s$ induced by our ignorance of the aforementioned neutrino parameters.  We devote special attention to theoretically well-motivated inflationary scenarios whose predictions for $\ns$ may be in perfect agreement with current observations, once uncertainties in the neutrino sector of the theory are properly included in the data analyses. 

The structure of the paper is as follows: we present the analysis method and datasets employed in this work in Sec.~\ref{sec:method}; results from current data are reported in Sec.~\ref{sec:current}, while Sec.~\ref{sec:4cast} is devoted to forecasts for future CMB experiments; implications of the results for inflationary models are summarized in Sec.~\ref{sec:inflation}. We draw our conclusions in Sec.~\ref{sec:conclusion}.  Our most important results may be found in Tables \ref{tab:ns_lcdmr}-\ref{tab:ns_lcdmrnnu} and Figures \ref{fig:2d_Nat}-\ref{fig:2d_4c_Higgs}.

\section{Method, Description of Cases Considered, and Data Sets}\label{sec:method}
In this section, we provide a detailed description of the statistical tools and datasets employed for drawing our results.
We also define the various cases we are treating with regards to a variety of possibilities for the spectrum of neutrino masses, the total 
neutrino mass, and $\Neff$. 

\subsection{Parameters}
We perform a Bayesian Monte Carlo Markov Chain analysis by making use of the latest version of the publicly available \texttt{cosmomc} package~\cite{Lewis:2002ah,Lewis:2013hha}, monitoring convergence through the Gelman and Rubin $R-1$ statistics~\cite{An98stephenbrooks}. Our baseline parameter vector is composed by the six $\lcdm$ parameters: the physical baryon density $\Omega_b h^2$, the physical cold dark matter density $\Omega_c h^2$, the angular size of the acoustic horizon at recombination $\theta$, the reionization optical depth $\tau$, the scalar spectral index $\ns$ and the logarithmic amplitude $\ln\left(10^{10} A_s\right)$ of the power spectrum of scalar perturbations at the pivot scale $k_p=0.05\,\mathrm{Mpc^{-1}}$. 

In addition to this standard set of parameters, we also consider (not necessarily at the same time) the tensor-to-scalar ratio $r$ (measured at the same pivot scale as the scalar perturbations), the total neutrino mass 
\begin{equation}
M_\nu =\sum_i m_i
\end{equation}
(where the index $i = 1,\,2,\,3$ runs over the three neutrino mass eigenstates), the neutrino hierarchy,
 and the number of relativistic degrees of freedom $\Neff$. In the {\bf baseline model}, these additional parameters are taken to be $r=0$, $M_\nu=0.059\eV$~\footnote{This is the lowest value allowed by neutrino oscillation data~(see e.g.~\cite{Forero:2014bxa,nufit}), which assumes a vanishing mass for the lightest mass eigenstate and normal neutrino mass hierarchy, and  it is usually approximated by $M_\nu=0.06\eV$  in different cosmological analyses (for example, those carried out by the Planck collaboration).} and $\Neff=3.046$. The inclusion of massive neutrinos is performed by considering different hierarchical scenarios and also different priors on the total neutrino mass, in order to test their possible impact on the inflationary observables. 

\subsection{Cases Considered}
We will compare the regions in the $n_s/r$ plane obtained by using the
standard approximate assumptions used throughout the literature in contrast with the regions obtained by using correct information
about neutrino masses from oscillations data.
We first consider the two standard approximate assumptions of:

\noindent
i) {\bf ``1+2'' case}:
\hfill\break
a single massive eigenstate with mass $M_\nu=0.059\,\eV$, plus two massless eigenstates, when the total mass is fixed 
\hfill\break
ii) {\bf ``3deg'' case}:
\hfill\break
fully degenerate scenario of three massive eigenstates with mass $M_\nu/3$ each, when the total mass is allowed to vary freely, provided $M_\nu\ge0$ .
\hfill\break
These two approximations mimic what is done in the Planck papers (and usually in the literature) when models with fixed or varying neutrino mass are analyzed (apart from the slightly different value of $M_\nu$ in the fixed mass case, see footnote 3).
These two cases are ``unphysical'' in that they do not match the correct values of neutrino mass as determined by oscillations data, 
yet up to now they have served as
reasonable approximations.  The question is at which point the data will become so good that these approximations are no longer adequate.

Then we compare the results arising from these approximate parameterizations with those obtained by implementing the exact neutrino mass hierarchy, either in the normal (labeled as ``{\bf NH}'') or inverted (``{\bf IH}'') hierarchy scenarios. From neutrino oscillation data, $m_1$ and $m_2$ are the eigenstates which are closer in mass, while the sign of $m^2_3-m^2_1$ determines the neutrino mass ordering (NH vs IH case). Thus, when using this ``exact'' parameterization, we sample over the lightest eigenstate ($m_1$ for NH or $m_3$ for IH), instead of sampling over  $M_\nu$, and compute the mass carried by the remaining eigenstates by means of the mass squared differences $\Delta m^2_{ij}\equiv m_i^2-m_j^2$ measured by oscillation experiments. In particular, we use the results of the global fit reported in Ref.~\cite{Forero:2014bxa}. Notice that, with this exact parameterization for the neutrino mass eigenstates, at least two out of the three of them are massive. As a result, when marginalizing over the total mass, the prior naturally imposed by oscillation measurements is $M_\nu\ge M_{\nu,\mathrm{min}}$, where $M_{\nu,\mathrm{min}}=0.059\,\eV$ ($M_{\nu,\mathrm{min}}=0.098\,\eV$) in the NH (IH) case. This is different from the prior imposed in the approximate case and this difference should be kept in mind when comparing results from the two parameterizations in the following sections, because of possible volume effects, on which we shall comment on later. 

For what concerns the effective number of relativistic species, when sampling over $\Neff$, we firstly impose a broad flat prior in the range $0 \le \Neff \le 10$. In addition, we also report results when considering the case of a hard prior $0\le \Neff\leq3.046$, as expected for example in low-reheating scenarios~\cite{Kawasaki:1999na,Kawasaki:2000en,Giudice:2000ex,Giudice:2000dp}, where the reheating temperature $T_\mathrm{RH}$ can be as low as $\mathcal{O}$(MeV). In both cases of the broad and hard prior on $\Neff$, we treat the difference $\Delta\Neff=\Neff-3.046$ between the value of $\Neff$ into consideration and the standard expected value of $3.046$ in the following way: $\Delta\Neff>0$ is considered as a massless neutrino contribution, \textit{i.e.} as an ``extra-radiation'' component; when $\Delta\Neff<0$, we instead rescale the (three) active neutrino temperature accordingly to $\Neff$, \textit{i.e.} the neutrino number density is rescaled by a factor $(\Neff/3.046)^{3/4}$. The neutrino density $\Omega_\nu h^2$ for a given total mass is then rescaled by the same factor.

\subsection{Datasets}
As our baseline dataset, we employ the full Planck 2015 measurements of the CMB anisotropies in temperature complemented with large-scale polarization~\cite{Aghanim:2015xee} (we refer to this combination as ``Planck TT+lowP''). We conservatively avoid use of small-scale polarization, since it could be still affected by a small amount of residual systematics~\cite{Ade:2015xua}. We also combine Planck data with the most recent degree-scale measurements of the BB power spectrum from the BICEP/Keck collaboration~\cite{Array:2015xqh} (BK14) and with geometrical BAO information from the galaxy surveys BOSS-DR11~\cite{Anderson:2013zyy}, the 6dF~\cite{Beutler:2011hx} and the MGS~\cite{Ross:2014qpa}.

In addition to deriving parameter estimates from current cosmological data, we also perform forecasts for future CMB experiments. We consider a future CMB satellite mission such as COrE+~\cite{core} and a future Stage IV (S4 hereafter) ground-based experiment (see e.g. Refs.~\cite{Abazajian:2013oma,Abazajian:2013vfg,Abazajian:2016yjj} for a summary of the expected performance in terms of parameter constraints). The mock data used consist in lensed temperature and polarization power spectra, generated according to Refs.~\cite{Bond:1998qg,Bond:1997wr}. We assume multifrequency coverage which allows perfect foreground removal and exquisite control of systematics. Specifications of the observed sky fraction, multipole coverage, beam width and sensitivity for the computation of noise spectra are set in agreement with \cite{Errard:2015cxa,rubmart} and references therein. Further details about experimental setup and the adopted fiducial model are provided in Sec.\ref{sec:4cast}.
We use a  prior on the reionization optical depth $\tau=0.06\pm0.01$ in combination with S4. We follow an exact likelihood formalism for the subsequent Monte Carlo analysis~\cite{Lewis:2005tp,Perotto:2006rj}.

\section{Present-day cosmological analyses}\label{sec:current}
In this section, we discuss the impact of massive neutrino properties, namely the total neutrino mass, its hierarchical distribution among massive neutrino eigenstates and the effective number of relativistic degrees of freedom at recombination, on the recovered value of the scalar spectral index $\ns$, in light of current cosmological data. 

We firstly focus on the comparison between the results arising from the baseline model ($\lcdm$) and its one-parameter extension $\lcdm+M_\nu$, provided $r=0$. We then perform an analogous comparison allowing also for a non-vanishing tensor component, parameterized via the tensor-to-scalar ratio $r$, to vary freely. Notice that in these cases we fix $\Neff$ to the standard value of 3.046. 

We shall discuss separately the effect of relaxing our assumptions about $\Neff$. In analogy to the cases depicted above, we shall thus compare the results from $\lcdm+\Neff$ and $\lcdm+\Neff+M_\nu$ fixing $r=0$. We then move to investigate the impact of varying the tensor-to-scalar ratio.

\subsection{Massive neutrinos with a vanishing tensor component}\label{subs:mnu}
As anticipated above, in this section we will fix the tensor-to-scalar ratio to zero and compare results coming from the baseline model and its one-parameter extension $\lcdm+M_\nu$.
We report our results on the scalar spectral index $\ns$ in Tabs. \ref{tab:ns_lcdm} and \ref{tab:ns_lcdmr} in terms of the $68\%$~CL intervals around the mean of the posterior distribution. Figure \ref{fig:1dns} visually summarises these results, in addition to a few more cases discussed in the following. We notice an overall agreement of the constraints on $\ns$ when different models and/or datasets are taken into account. However, small departures \footnote{When we assess the magnitude of the shift between two given values of the spectral index with mean value $n_{s,i}$ and $1\sigma$ error $\sigma_i$ (with $i=1,2$) in units of $\sigma$, we quote the following quantity: $|n_{s,1}-n_{s,2}|/\sqrt{\sigma_1^2+\sigma_2^2}$. In the case when the 68\% CL is not symmetric around the mean, we take the half width of the same range as an estimate for $\sigma_i$.}  -- at the level of a few fractions of $\sigma$-- from the mean value obtained for the $\lcdm$ model are observed, which can be explained in terms of physical effects of the neutrino background on cosmological observables, on which we shall comment thouroughly in this section.

The aim of this section is threefold: namely, we want i) to study the effect of the neutrino mass splittings as well as ii) to assess the impact of the marginalization over the total neutrino mass (as opposed to fixing the mass to a given value), focusing mostly on the estimates of the scalar spectral index $n_s$, and iii) provide a thorough physical explanation underlying the observed shifts in the scalar spectral index. In this section we will then focus on the standard $\Lambda$CDM scenario (with a total neutrino mass fixed to $M_\nu = 0.059\,\eV$) and on its one-parameter extension, in which $M_\nu$ is allowed to vary freely and marginalized over, dubbed as $\Lambda$CDM+$M_\nu$. In the context of bayesian analysis, the marginalization over one parameter allows to take into account any possible effect whose imprecise knowledge could have on the determination of the remaining parameters of the model. These effects, such as bias in the recovered mean values and/or broadening of the confidence levels, may be otherwise hidden by fixing the unknown parameter to a specific value. 

\subsubsection{Neutrino Mass Splitting}
We shall consider different parameterizations for the splittings of the total neutrino mass among the mass eigenstates. In particular, we compare results considering the exact mass distribution according to the NH scenario on one side, with those arising from the usual approximations. These approximations consist in: either a single massive eigenstate carrying the total mass, fixed to the minimum value allowed by neutrino oscillation measurements, and two massless eigenstates (``1+2'' scenario, for $\lcdm$ and generically all those models where the total neutrino mass is fixed to the minimum value allowed by oscillations), or three fully degenerate massive neutrinos (``3deg'', for $\lcdm+M_\nu$ and generically all those models where the total neutrino mass is allowed to vary freely and not fixed to any specific value), on the other side.

For the sake of comparison, we also report results derived from assuming an exact distribution according to the inverted hierarchy (IH) scenario for some specific models analyzed in this work. However, we choose to mainly focus on the NH case, which seems to be slightly preferred by current cosmological limits on $M_\nu$ (a combination of cosmological measurements is close to disfavour $M_\nu>M_{\nu,\mathrm{min}}$, where $M_{\nu,\mathrm{min}}=0.098\,\eV$ is the minimal mass allowed by oscillation measurements in the inverted hierarchy scenario, at $\sim2\sigma$ \cite{Palanque-Delabrouille:2014jca,Cuesta:2015iho,Giusarma:2016phn,Hannestad:2016fog,
Alam:2016hwk,DiValentino:2015sam}) and very mildly preferred by the latest neutrino oscillation data~\cite{nova}.

In comparing the results arising from the different assumptions listed above, we would like to highlight possible deviations in the recovered mean value of $\ns$ due to a different neutrino scenario, possibly hinting at some sensitivity from current cosmological data to the neutrino mass splittings.

In Tab.~\ref{tab:ns_lcdm}, we compare constraints on $\ns$ obtained in the context of the $\lcdm$ model, with the total neutrino mass fixed to the minimum value allowed by oscillations, \textit{i.e.} $M_\nu=0.059\,\eV$, with those obtained for the $\lcdm+M_\nu$ model, after marginalization over $M_\nu$. 
We test any model against the Planck TT+lowP data alone and in combination with BAO. 

Table ~\ref{tab:ns_lcdm} also contains information about the comparison between different mass splittings, labeled as NH for normal hierarchy and ``approx'' for either ``1+2'' or ``3deg''. We shall focus on this comparison firstly. Notice that, for each model and dataset combination, the two mass parameterizations do not provide precisely the same constraints. Focusing on the $\lcdm$ scenario, the shift in the mean value of $n_s$ is negligible and we cannot exclude a statistical fluctuation of the MCMC analysis as a valid explanation. However, we notice that for the two combinations of data sets considered, the shift is going in the same direction, \textit{i.e.} lowering the value of $\ns$ when the NH parameterization is used. 

\begin{figure}
\begin{center}
\includegraphics[width=0.9\columnwidth]{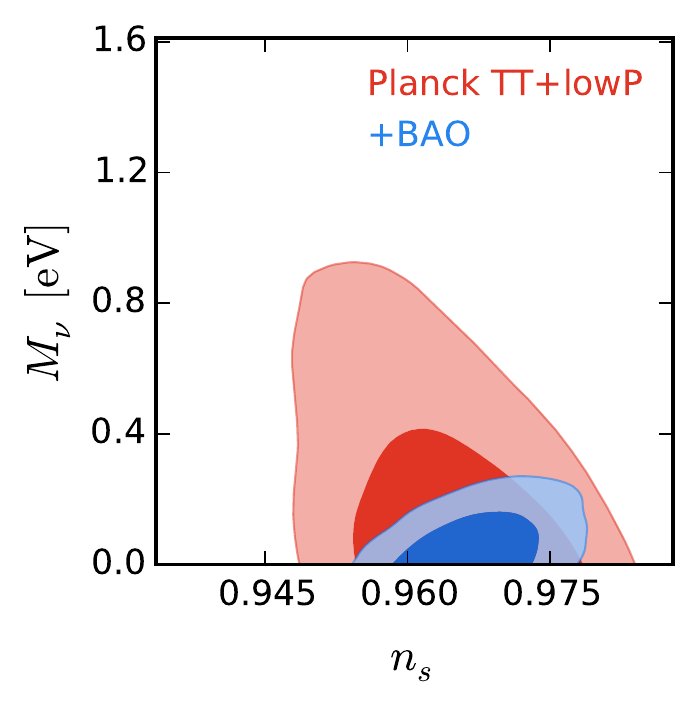}
\end{center}
\caption{Two-dimensional probability contours at 68\% and 95\% CL in the $\ns-M_\nu$ plane for the Planck TT+lowP dataset and the $\Lambda\mathrm{CDM}+M_\nu$ model, when assuming three fully degenerate massive neutrinos (labeled as ``3deg'' scenario in the main text).}\label{fig:2dnsmnu}
\end{figure}

In the $\lcdm+M_\nu$ scenario, the shift in $\ns$ between the two parameterizations is more pronounced, albeit it is still small, at the level of $0.1\sigma$. Interestingly, the direction of the shift is opposite for this model if one tests it against different combination of datasets. In particular, considering Planck TT+lowP alone, $\ns$ increases going from NH to ``3deg'' (compare first and second rows in the $\lcdm+M_\nu$ column of Tab.\ref{tab:ns_lcdm}). In contrast, $\ns$ decreases going from NH to ``3deg'', when the combination of Planck TT+lowP+BAO is employed (compare third and fourth rows in the $\lcdm+M_\nu$ column of Tab.\ref{tab:ns_lcdm}).
In both cases, the nature of the shift is found in the correlation \footnote{We shall talk equivalently about correlation or degeneracy when referring to the fact that any change to one parameter induces a modification to another parameter.} arising between $\ns$ and $M_\nu$, shown in Fig.~\ref{fig:2dnsmnu} for the ``3deg'' parameterization (NH provides similar contours) and discussed below. 

Concerning CMB measurements alone, the increase in $\ns$ when moving from NH to ``3deg''  arises from the fact that the power in the damping tail can be kept approximately constant by increasing $M_\nu$ (enhancing power in the tail by suppressing structures and hence the lensing potential) and decreasing $n_s$ (thus tilting the spectrum to give less power to the small scales). Since the posterior distribution for $M_\nu$ starts from zero in the ``3deg'' case, while for the NH scenario values below $0.059\,\eV$ are not allowed for $M_\nu$, the center of mass of the posterior distribution for $M_\nu$ is shifted to larger values in the latter case with respect to the former. Given the correlation between $M_\nu$ and $\ns$ discussed just above, this yields a smaller value for $\ns$ in the NH case. In fact, we have checked that we are able to reduce significantly the shift if we impose a hard prior of $M_\nu>0.059\eV$ also in the ``3deg'' case, making clear that this is mainly a volume effect. 


Notice that $\ns$ and $M_\nu$ are anti-correlated (\textit{i.e.} higher values of $M_\nu$ correspond to lower values of $\ns$) when Planck TT+lowP data are used. In contrast, the two parameters are positively correlated when BAO information are added to CMB data, as clearly visible in Fig.~\ref{fig:2dnsmnu}. As an example, the correlation coefficient defined as $R=C_{ij}/\sqrt{C_{ii}C_{jj}}$, where $C$ is the covariance matrix of cosmological parameters and $i,j=\ns,M_\nu$, changes from $R=-0.45$ (implying negative correlation) for Planck TT+lowP to $R=0.34$ (implying positive correlation) when BAO measurements are added \footnote{These numbers refer to the ``3deg'' case. Similar figures apply to the NH case.}. 
 


An explanation for the aforementioned degeneracies when BAO data are added can be found by studying the physical effect of neutrino masses on the quantities constrained by BAOs. Recall BAO data constrain the ratio $D_v/r_s(z_{\text{drag}})$, where $r_s(z_{\text{drag}})$ denotes the sound horizon at the drag epoch (\textit{i.e.} the epoch at which baryons decouple from photons, slightly after recombination, when photon pressure is no longer available to prevent gravitational instability), and $D_v$ is a distance combination. In particular, $D_v$ is a combination of the line-of-sight comoving distance related to the Hubble parameter $H$, and the transverse comoving distance $D_M$ \footnote{The transverse comoving distance $D_M$ is related to the angular diameter distance $D_A$ via the relation $D_M=(1+z)D_A$.}:
\begin{eqnarray}
D_v(z) = \left [ D_M(z)^2\frac{cz}{H(z)} \right ]^{\frac{1}{3}} \, ,
\end{eqnarray}
which BAO measurements are mostly sensitive to in angle-averaged statistics. As $M_{\nu}$ is increased keeping $\Omega_bh^2$ and $\Omega_ch^2$ fixed, the early-time expansion rate increases and hence in order to keep $\theta$ fixed (which controls the scale of the first peak), $\Omega_{\Lambda}$ must decrease. As $\Omega_{\Lambda}$ decreases, $D_v(z)$ increases and correspondingly both $r_s/D_v$ and $H_0$ decrease. This behaviour explains why BAO data, by excluding lower (higher) values of $H_0$ ($\Omega_mh^2$), exclude the region associated to higher $M_{\nu}$ and correspondingly, prefer higher values of $n_s$. In this way, the anti-correlation between $M_{\nu}$ and $n_s$ present when CMB data alone is employed, is reverted. Further discussions on these effects can be found in~\cite{Hou:2012xq}.

\subsubsection{Total Neutrino Mass}
We will now focus on the comparison between the recovered values of $\ns$ for a given hierarchy, \textit{i.e.} we discuss possible deviations due to a different choice of the cosmological model (either $\lcdm$ or the one-parameter extension $\lcdm+M_\nu$) and/or dataset combination.

Marginalizing over the total neutrino mass introduces shifts in $\ns$ with respect to the $\lcdm$ model, meaning that the unknown value of the total neutrino mass may play a non-negligible role in recovering the exact constraints on the scalar spectral index. The shift in $\ns$ due to the marginalization over $M_\nu$ goes in the direction of lowering $\ns$ if the models are tested against CMB only ($\sim0.3\sigma$ in the NH case and $\sim0.2\sigma$ in the ``3deg'' case, with respect to the $\lcdm$ model), while it goes in the opposite direction when BAO data are also considered ($\sim0.2\sigma$ in the NH case and $\sim0.1\sigma$ in the ``3deg'' case, with respect to the $\lcdm$ model). 

\begin{figure}
\begin{center}
\includegraphics[width=0.99\columnwidth]{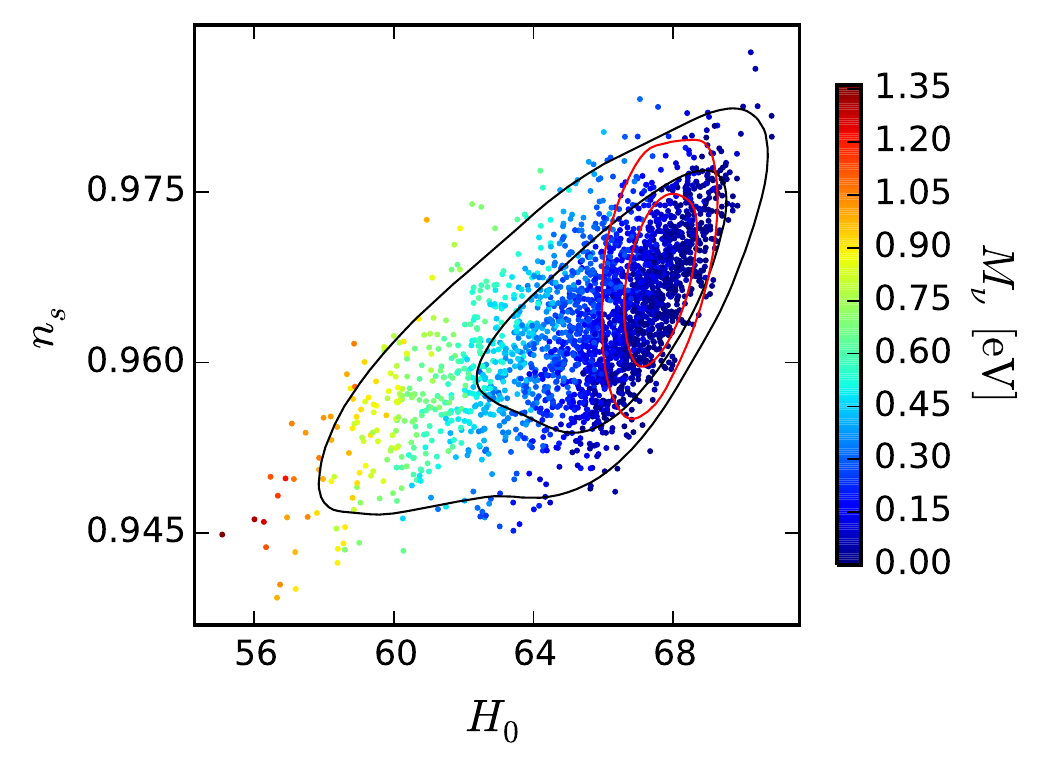}
\end{center}\caption{Colored scatter plot showing the mutual degeneracies between $\ns$, $H_0$ (in $\mathrm{km\,s^{-1}\,Mpc^{-1}}$) and $M_\nu$ (in $\eV$) for the $\lcdm+M_\nu$ model in the ``3deg'' case, \textit{i.e.} a cosmological scenario with three massive and fully degenerate neutrinos. Each point represents a model in the $\ns/H_0$ plane. The color code represents different neutrino masses, according to the vertical bar on the right of the figure. The two-dimensional contours are for Planck TT+lowP (black) and Planck TT+lowP+BAO (red). The inclusion of BAO clearly excludes the high-mass region (red and yellow points) of the parameter space.}\label{fig:scatter}
\end{figure}

\begin{figure*}
\begin{center}
\includegraphics[width=0.9\textwidth]{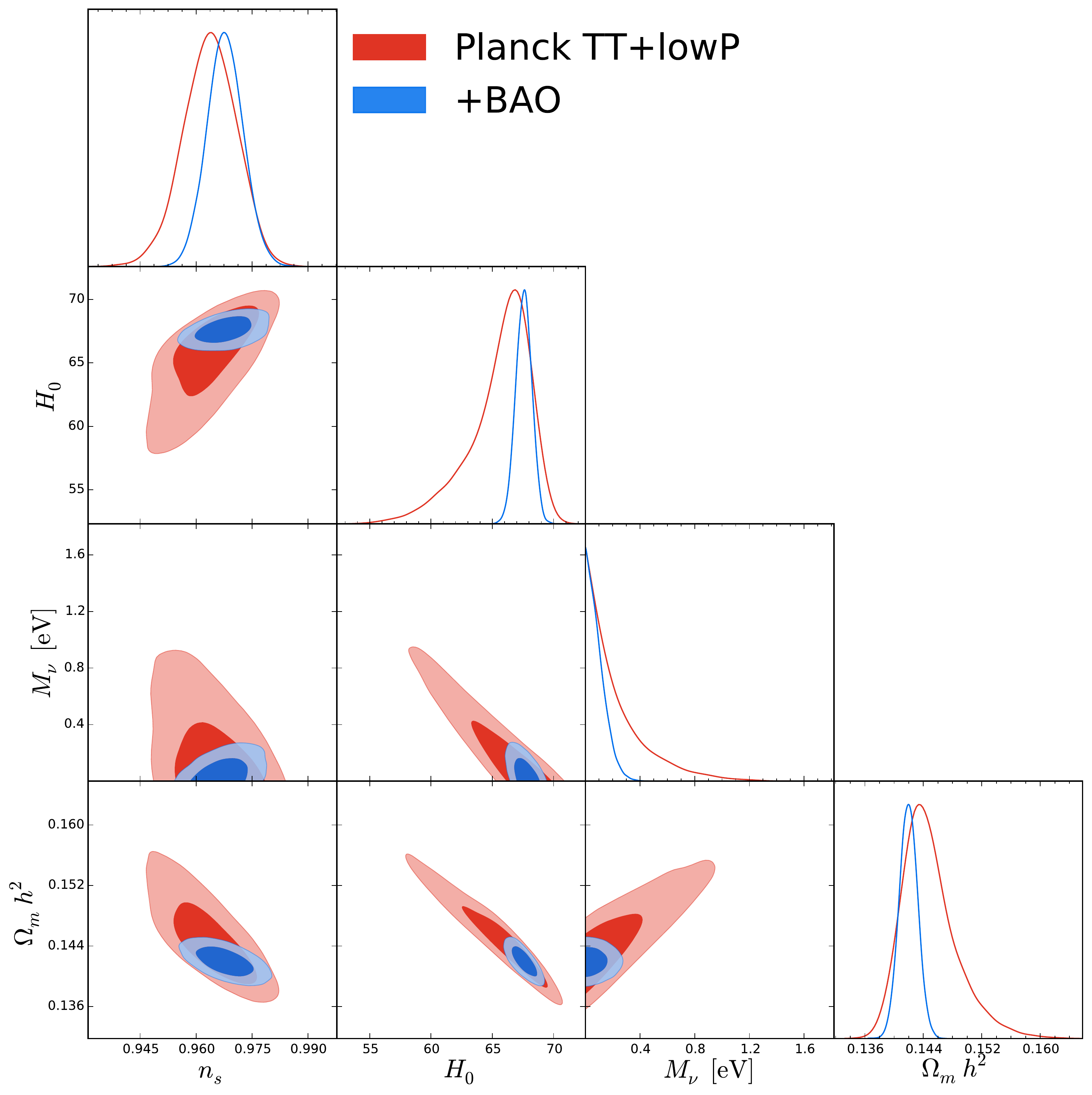}
\end{center}
\caption{Two-dimensional probability contours at 68\% and 95\% CL, and one-dimensional posterior probability distributions showing the main correlations among cosmological parameters responsible for the shift in $\ns$, for the indicated dataset and the $\lcdm+M_\nu$ model, when assuming three fully degenerate massive neutrinos (labeled as ``3deg'' scenario in the main text).}\label{fig:tri}
\end{figure*}

We have already seen that the addition of BAO data is, in general, responsible for an increase in $\ns$, quantified in $\sim0.2\sigma$ in the $\lcdm$ scenario and $\sim0.5\sigma$ in the $\lcdm+M_\nu$ scenario, with respect to the equivalent values obtained with the Planck TT+lowP datasets only, almost independently on the choice of the mass splitting. The reason for these shifts, extensively discussed above, is related to degeneracies arising between $\ns$, the Hubble constant $H_0$ and the matter density $\Omega_mh^2$, shown in Fig.~\ref{fig:tri} for the $\lcdm+M_\nu$ model (a similar figure is obtained for the $\lcdm$ model). BAO data are able to exclude lower (higher) values of $H_0$ ($\Omega_mh^2$), thus reducing the volume of the parameter space corresponding to the low $\ns$ region (see also Fig.\ref{fig:scatter}, where the two-dimensional probability contours in the $\ns-H_0$ plane are colored with respect to the allowed value of $M_\nu$). BAO measurements are able to measure $\Omega_m$ and, in combination with CMB, are able to measure $H_0$, so they essentially split $\Omega_m$ and $H_0$ (see e.g.~\cite{2013PhR...530...87W, Addison:2013haa}).

The effect of adding BAO is also clearly visible in Fig.~\ref{fig:2dnsmnu}, where the direction of the correlation in the $\ns-M_\nu$ plane found by combining CMB and BAO (blue contours) is almost orthogonal to the direction identified with CMB alone. The impact of a free neutrino mass on $\ns$, \textit{i.e.} the increase of the mean value of the scalar spectral index with respect to the $\lcdm$ model, is less pronounced in the ``3deg'' case than in the NH scenario. In fact, having access to the $M_\nu<0.059\eV$ region of the parameter space mitigates the effect of marginalizing over the neutrino mass.



For completeness, we shall report here the 68\% bounds on $\ns$ in the case of an IH neutrino mass spectrum when the neutrino mass is marginalized over. In this case, we get $\ns=0.9627^{+0.0074}_{-0.0066}$ for Planck TT+lowP and $n_s=0.9693\pm 0.0046$ when BAO measurements are also included. These constraints are perfectly in agreement with the picture depicted above. In fact, we obtain a further reduction of $\ns$ with respect to the ``3deg'' case when only CMB data is exploited, while we observe a further increase in the value of $n_s$ when BAO measurements are added. This behaviour is again due to the reduced probability volume available in the IH regime when $M_\nu$ is allowed to vary freely. In fact, the IH scenario implies $M_\nu>0.098\eV$, \textit{i.e.} the lowest mass value allowed by oscillation measurements once the IH scheme is assumed, to be compared with the equivalent priors on $M_\nu$ in the NH case ($M_\nu>0.059\,\eV$) and ``3deg'' case ($M_\nu>0\,\eV$), thus enhancing the volume effects already discussed in this section. 

\begin{table}[h!]
\begin{tabular}{cc||c|c|}
&&$\Lambda\mathrm{CDM}$	&$\Lambda\mathrm{CDM}+M_\nu$\\
\hline
\multirow{2}{*}{Planck TT+lowP} &NH &$ 0.9655\pm 0.0063$	& $0.9629\pm 0.0069$\\	
                                                    &approx	& $ 0.9656\pm 0.0063$		& $0.9636\pm 0.0071$\\
\hline
\multirow{2}{*}{+BAO} &NH & $0.9671\pm 0.0045$	& $0.9686\pm 0.0047$\\	
                                    &approx	& $ 0.9673\pm 0.0045$		& $0.9678\pm 0.0048$\\
\end{tabular}
\caption{68\% probability intervals around the mean for the scalar spectral index $\ns$ for the indicated datasets and cosmological models. The lines labeled as ``approx'' refer to the ``1+2'' (first column, $\lcdm$) and ``3deg'' (second column, $\lcdm+M_\nu$) parameterizations used when either $M_\nu$ is fixed or marginalized over, respectively. The two columns refer to the following two cases:  (i) the first column, dubbed as standard $\lcdm$ scenario has a total neutrino mass fixed to $M_\nu=0.059\,\eV$ and  (ii) the second column, dubbed as $\lcdm+M_\nu$ is for a one-parameter extension in which $M_\nu$ is free to vary ($M_\nu\ge0$ for ``approx" i.e. ``3deg" case and $M_\nu\ge 0.059$ eV for the NH case).}\label{tab:ns_lcdm}
\end{table}

In conclusion, in the context of the $\lcdm$ model, with a vanishing tensor component, the effect on $\ns$ of an exact modelling of the neutrino mass splitting, either NH or IH, as opposed to an approximate parameterization ``1+2'' is tiny, safely allowing to make use of the approximate parameterization instead of the exact spectrum when the neutrino mass is fixed. The main impact on the scalar spectral index is induced by relaxing the assumption about the total mass. Indeed, if one marginalizes over the total neutrino mass, the constraints on $\ns$ are shifted and broadened with respect to the values obtained in the baseline model. In this context, the choice of the hierarchy, either exact or approximate, could play a role, mainly due to the different prior on $M_\nu$ adopted in the two cases ($M_\nu>M_{\nu,\mathrm{min}}$, with $M_{\nu,\mathrm{min}}\neq0$ in the exact parameterization)\footnote{Cosmological models predicting a vanishing neutrino density today have been proposed~\cite{Beacom:2004yd}, These models can motivate, from a theoretical point of view, the choice of a vanishing lower cutoff $M_{\nu,\mathrm{min}}=0$, seen as a phenomenological proxy of the effect of a smaller density of neutrinos with respect to the expectation in standard $\lcdm$.}.

\subsection{Massive neutrinos with a non-vanishing tensor component}\label{subs:mnu2}

The impact of massive neutrino properties on the constraints on the scalar spectral index might be relevant when assessing the agreement of the predictions from different inflationary models with observations, \textit{i.e.} when exploring the $\ns/r$ plane, where $r$ is the tensor-to-scalar ratio. For this reason, we also include here a very simple extension of the minimal $\lcdm$ scheme, namely the $\lcdm+r$ model. In analogy to what we have presented before, we report our results in Tab.~\ref{tab:ns_lcdmr}, which are equivalent to those presented in Tab.~\ref{tab:ns_lcdm}, but with the addition of BAO measurements, along with BK14 data.

By comparing the results reported in Tab.~\ref{tab:ns_lcdm} with those reported in Tab.~\ref{tab:ns_lcdmr}, it is evident that the marginalization over $r$ has the overall effect of increasing the mean value of $\ns$ for each of the aforementioned parameterizations and data combinations. As a reference, for the $\lcdm+r$ model when the total neutrino mass is fixed to $M_\nu=0.059\eV$ in the NH case and the Planck TT+lowP dataset is used, the mean value of $\ns$ is shifted by a factor of $\sim0.1\sigma$ with respect to the $\lcdm$ scenario (\textit{i.e.} when $r=0$). This feature, \textit{i.e.} the increase in $\ns$ when $r$ is non-vanishing, is expected, since the tensor contribution adds power to the TT spectrum at large scales. This effect can be compensated by a higher value of $\ns$. 

This mild correlation between the two parameters can be broken by the inclusion of direct measurements of the BB spectrum, provided by BK14. Indeed, the inclusion of BK14 data, being able to further reduce the upper bound on $r$, has the effect of lowering the mean value of $\ns$ with respect to the results obtained from Planck TT+lowP. For comparison, the $\lcdm+r$ model, when $M_\nu=0.059\eV$ in the NH case, tested against Planck TT+lowP provides $\ns=0.9666\pm0.0062$ at 68\% CL, which decreases down to $n_s=0.9656\pm 0.0062$ when BK14 is added (a value in perfect agreement with the results arising from the $\lcdm$ model and Planck TT+lowP analyses). 

We will now focus on Tab.~\ref{tab:ns_lcdmr}. Firstly, let us compare the results from the different neutrino mass splittings \textit{at fixed total neutrino mass}.
The shifts in the recovered value of $\ns$ when considering either the NH or the ``1+2'' cases are not very significant. Still, it is interesting to understand why the mean value of $n_s$ is lowered when
moving  from the NH to the ``1+2'' parameterization in the $\lcdm+r$ model, whereas it increases
when the same move is made in  the $\lcdm$ scenario (see Tab.~\ref{tab:ns_lcdm}). We 
note that the move from NH to ``1+2" essentially amounts to adding an additional massive neutrino (from one to two).  To understand the observed trend, we have performed an additional MCMC run with the following choice for the mass splitting: we have assumed the ``NH'' parameterization and fixed the total mass to $M_\nu=0.07\eV$, slightly higher than $M_{\nu,\mathrm{min}}$ in the NH scenario. With this choice, the mass of the lightest eigenstate has been fixed to a non-vanishing but still small value, so that all the three eigenstates are massive (recall that, for $M_\nu=0.059\,\eV$, we have two massive neutrinos in the NH case and only one massive neutrino in the ``2+1'' case). 

\begin{figure}
\begin{center}
\includegraphics[width=1\columnwidth]{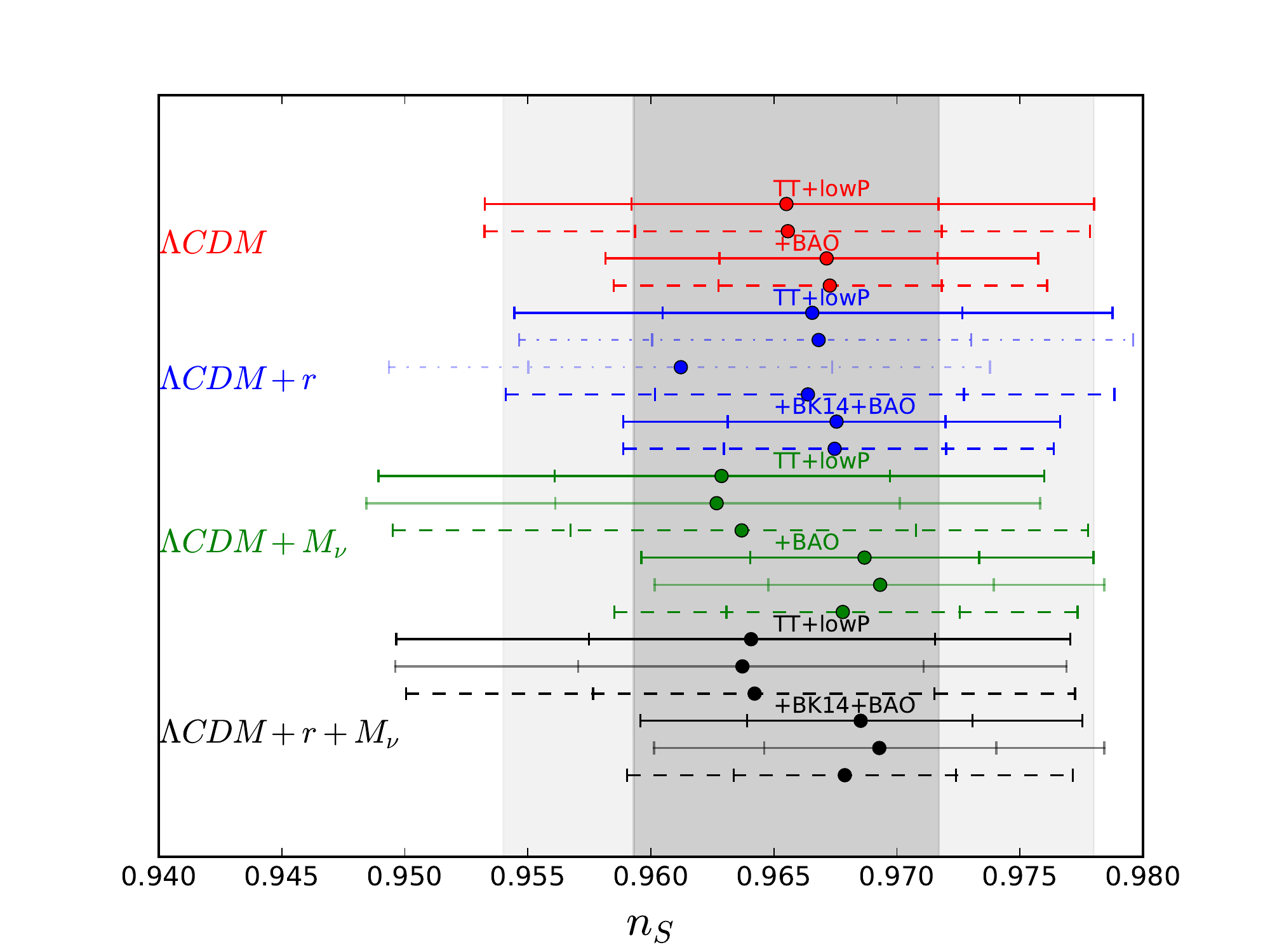}
\end{center}
\caption{Marginalized confidence intervals for the scalar spectral index $\ns$ for the indicated cosmological models and datasets. Solid bold lines are for the exact NH parameterization (total neutrino mass distributed according normal hierarchical scenario), solid light lines for the IH parameterization (total neutrino mass distributed according inverted hierarchical scenario), dashed lines for the degenerate parameterization (one massive neutrino plus two massless species when $M_\nu$ is fixed, three degenerate massive neutrinos when $M_\nu$ is allowed to vary). The two dashed-dotted blue lines are drawn for the NH parameterization when fixing the lightest eigenstate to $m_1=0.009\,\eV$ (first dashed-dotted line from the top) and $m_1=0.164\eV$ (second dashed-dotted blue line from the top), respectively. The vertical bands are 68\% and 95\% ~CL limits from Planck TT+lowP in the context of a $\lcdm$ model with one massive neutrino with mass $M_\nu=0.06\eV$.}\label{fig:1dns}
\end{figure}

The allowed interval for the spectral index when $M_\nu=0.07\,\eV$ is $\ns=0.9668\pm 0.0064$ at 68\% CL. This result is slightly higher than the corresponding value $\ns=0.9666\pm 0.0062$ for the $\lcdm+r$ model with $M_\nu=0.059\,\eV$ (see Fig.~\ref{fig:1dns}), \textit{i.e.} with the lightest eigenstate behaving as a fully relativistic species, and $\sim0.1\sigma$ higher than the value $\ns=0.9664\pm0.0063$ for the $\lcdm+r$ model and the ``1+2'' parameterization, \textit{i.e.} when two out of three neutrinos act as fully relativistic species.
Notice that $M_\nu=0.07\,\eV$ is almost equivalent to $M_\nu=0.059\,\eV$ from the point of view of background evolution, \textit{i.e.} the difference in energy density associated with the total neutrino mass is negligible. What is changing in the three cases listed above, namely NH with $M_\nu=0.07\,\eV$, NH with  $M_\nu=0.059\,\eV$ and ``2+1'' with $M_\nu=0.059\,\eV$, is the number of fully relativistic species. This might suggest that assuming the presence of species which remain relativistic up to the present time can play a role in constraining $\ns$. However, the significance of the shifts in $\ns$ previously reported is very mild. Furthermore, in the $\lcdm+r+M_\nu$ scenario one cannot identify a trend in the mean value of $\ns$ as neat as the one arising in the $\lcdm+r$ model, so we cannot exclude different explanations for the shifts in $\ns$ from the one reported above, such as the effect of degeneracies with other cosmological parameters and/or statistical fluctuations. 

We will now focus on the comparison between the results obtained in the two different cosmological models, $\lcdm+r$ and $\lcdm+r+M_\nu$, for the combination of dataset reported in Tab.~\ref{tab:ns_lcdmr}.
In analogy to what discussed in the previous section, the marginalization over the total neutrino mass has the overall effect of lowering the mean value of $\ns$ when CMB data only (Planck TT+lowP alone and/or in combination with BK14) are exploited. The amount of the shift is greater than $0.2\sigma$ with respect to the $\lcdm+r$ model. As a reference, focusing on the NH parameterization, we get a $\sim0.3\sigma$ shift for Planck TT+lowP and $\sim0.2\sigma$ shift for Planck TT+lowP+BK14. An additional test which confirms the inverse correlation between $\ns$ and $M_\nu$ has been performed by extracting the 68\% CL for $\ns$ in the $\lcdm+r$ model with NH parameterization and the total neutrino mass fixed to $M_\nu=0.5\eV$, value chosen to enlarge the impact of higher neutrino masses on the scalar spectral index. We recover $\ns=0.9612\pm 0.0062$, $\sim0.6\sigma$ lower than the corresponding value for the same model with $M_\nu=0.059\eV$.

The addition of BAO measurements inverts the correlation between $\ns$ and $M_\nu$, as detailed in the previous section, thus resulting in a higher mean value of the spectral index. We notice an increase in $\ns$ i) if we fix the cosmological model and compare CMB alone (Planck TT+lowP+BK14) with CMB+BAO (\textit{i.e.} if we compare either the first or second row with the last row of Tab.~\ref{tab:ns_lcdmr}; compare e.g. $\ns=0.9666\pm0.0062$ for Planck TT+lowP in the $\lcdm+r$ model and ``NH'' case with $\ns=0.9676\pm0.0045$ for Planck TT+lowP+BK14+BAO in the $\lcdm+r$ model and ``NH'' case) and ii) if we fix the dataset to be CMB+BAO and compare the two cosmological models (\textit{i.e.} if we focus on the last row of Tab.~\ref{tab:ns_lcdmr} and compare the two columns; compare e.g. $\ns=0.9676\pm0.0045$ in the $\lcdm+r$ model and ``NH'' case with $\ns=0.9677\pm0.0046$ in the $\lcdm+M_\nu$ model and ``NH'' case). 
We can derive that the shift observed in case i) is larger than that obtained in case ii). In addition, the shift induced by the marginalization over $M_\nu$ is less pronounced for Planck TT+lowP+BK14+BAO than for the other dataset combination. 

An illustrative summary of this section is provided in Fig.~\ref{fig:1dns}, which depicts the 68\% and 95\% CL allowed ranges for $\ns$ for the several combinations of datasets and for the different cosmological models analysed. 

Before concluding this section, we would like to emphasise an interesting finding. Notice that adding BK14 helps lower the mean value of $\ns$ when $M_\nu$ is fixed, as reported in Tab.~\ref{tab:ns_lcdmr}. The same does not happen when $M_\nu$ is marginalized over. A suitable explanation is that $M_\nu$ and $r$ are slightly degenerate for CMB only. Thus, the addition of direct measurements of the power spectrum of the B-modes of CMB polarization (BB power spectrum) provides better constraints on $r$ when $M_\nu$ is fixed, rather than marginalized over.

From what reported in this section, we can thus conclude that, in the presence of a non-vanishing tensor component ($r\neq0$), the recovered value of $\ns$ is stable against both the choice of the hierarchy and the value of the total neutrino mass \textit{if a combination of CMB and BAO data is used}. This implies a value of the spectral index of $\ns=0.9677\pm0.0046$ in the $\lcdm+r+M_\nu$ model, with NH parameterization, $\sim0.1\sigma$ lower than the corresponding value ($\ns=0.9686\pm0.0047$) for Planck TT+lowP+BAO and $r=0$. If otherwise only CMB measurements are employed, the same conclusions derived in the previous section about the role of the hierarchy and/or the exact value of the total neutrino mass also apply here. The recovered value of the spectral index in the $\lcdm+r+M_\nu$ model in the NH scenario tested against Planck TT+lowP+BK14 is in this case $\ns=0.9641\pm0.0064$.

\begin{table}[h!]
\begin{tabular}{cc||c|c|}
&&$\Lambda\mathrm{CDM}+r$	&$\Lambda\mathrm{CDM}+r+M_\nu$\\
\hline
\multirow{2}{*}{Planck TT+lowP} &NH &$0.9666\pm 0.0062$	& $0.9640^{+0.0075}_{-0.0066}$\\	
                  &approx	& $ 0.9664\pm 0.0063$		& $0.9642^{+0.0073}_{-0.0066}$\\
\hline
\multirow{2}{*}{+BK14} &NH & $0.9656\pm 0.0062$	& $0.9641\pm 0.0064$\\	
                  &approx	& $ 0.9654\pm 0.0062$		& $0.9640\pm 0.0066$\\
\hline
\multirow{2}{*}{+BAO} &NH & $0.9676\pm 0.0045$	& $0.9677\pm 0.0046$\\	
                  &approx	& $ 0.9675\pm 0.0045$		& $0.9679\pm 0.0046$\\
\end{tabular}
\caption{68\% probability intervals around the mean for the scalar spectral index $\ns$ for the indicated datasets and cosmological models. The lines labeled as ``approx'' refer to the ``1+2'' (first column, $\lcdm+r$) and ``3deg'' (second column, $\lcdm+r+M_\nu$) parameterizations used when either $M_\nu$ is fixed or marginalized over, respectively. The two columns refer to the following two cases:  (i) the first column, dubbed as $\lcdm+r$ scenario, has a total neutrino mass fixed to $M_\nu=0.059\,\eV$ and  (ii) the second column, dubbed as $\Lambda$CDM$+r+M_\nu$, is for a one-parameter extension with respect to the first column in which $M_\nu$ is free to vary, as described in Tab.~\ref{tab:ns_lcdm}. }\label{tab:ns_lcdmr}
\end{table}

\subsection{Relativistic degrees of freedom with a vanishing tensor component}\label{subs:nnu}
In this section, we discuss the impact that accounting for a number of relativistic degrees of freedom $\Neff$ different from the standard value of 3.046 might have on the recovered value of the scalar spectral index $\ns$. We will follow the same approach outlined for massive neutrinos. We firstly assume a vanishing tensor component ($r=0$) and compare results coming from two cosmological models: $\lcdm+\Neff$, where the total neutrino mass is fixed to the minimal value allowed by oscillations ($M_\nu=0.059\,\eV$), and $\lcdm+\Neff+M_\nu$, where we allow the total neutrino mass to vary freely. For the sake of simplicity, and taking into account the results of the previous section, \textit{i.e.} that the exact choice of the neutrino hierarchy plays a marginal role in modifying the constraints on $\ns$, we choose to follow here the approximate parameterization described in the previous section: we assume a ``1+2'' scenario when the total neutrino mass is fixed and a ``3deg'' scenario when it is varied. We will see indeed how the marginalization over $\Neff$ considerably broadens the constraints on $\ns$, thus hiding the tiny shifts possibly induced by the different choices of neutrino hierarchy.

For each model tested in this section, we consider two possible scenarios when varying $\Neff$. In one case, we make use of a broad prior on $\Neff$, allowing it to vary freely between 0 and 10. When doing so, we are not applying any prior information on the number of relativistic degrees of freedom and simply let the data tell us what is the preferred value within the context of the cosmological model under scrutiny; we refer to this scenario as ``broad''. In the other case, we instead apply a hard prior of $\Neff\leq3.046$ when varying $\Neff$. The aim of this choice is to test a scenario in which no extra-radiation is allowed, but the number of relativistic species could be lower than the standard expected value due to incomplete thermalization processes, as the case in very low-reheating scenarios; we shall refer to this case with the ``hard'' label.

\begin{figure*}
\begin{center}
\includegraphics[width=0.9\textwidth]{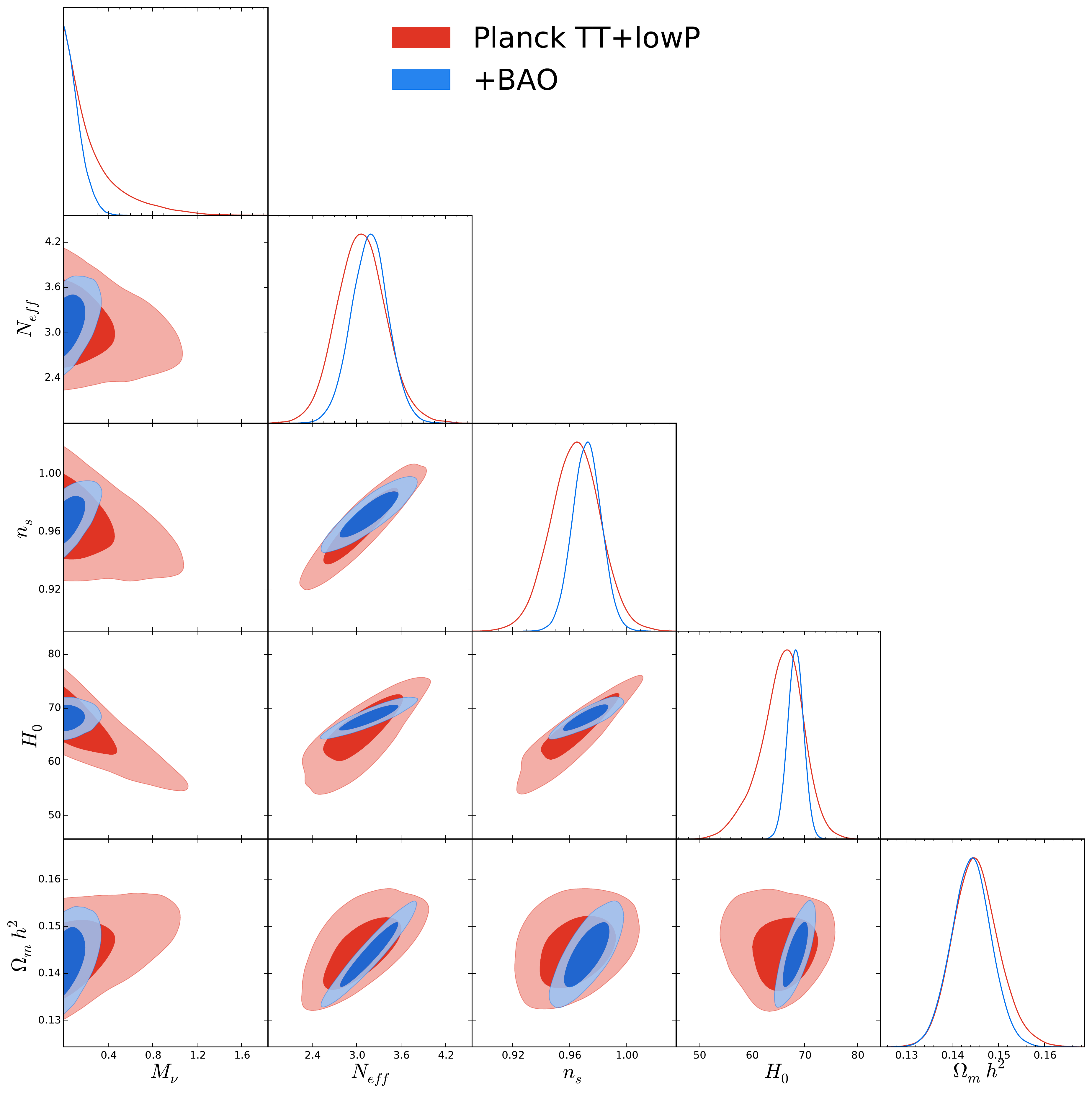}
\end{center}
\caption{Two-dimensional probability contours at 68\% and 95\% CL, and one-dimensional posterior probability distributions showing the main degeneracies among cosmological parameters responsible for the shift in $\ns$, for the indicated dataset and the $\lcdm+\Neff+M_\nu$ model. In plotting the figure, we have assumed three massive degenerate neutrinos (``3deg'' parameterization in the main text) and a broad prior on $\Neff$ ($0\leq\Neff\leq10$, see the text for further details about additional priors adopted on $\Neff$).}\label{fig:tri_nnu}
\end{figure*}

\begin{figure}
\begin{center}
\includegraphics[width=1\columnwidth]{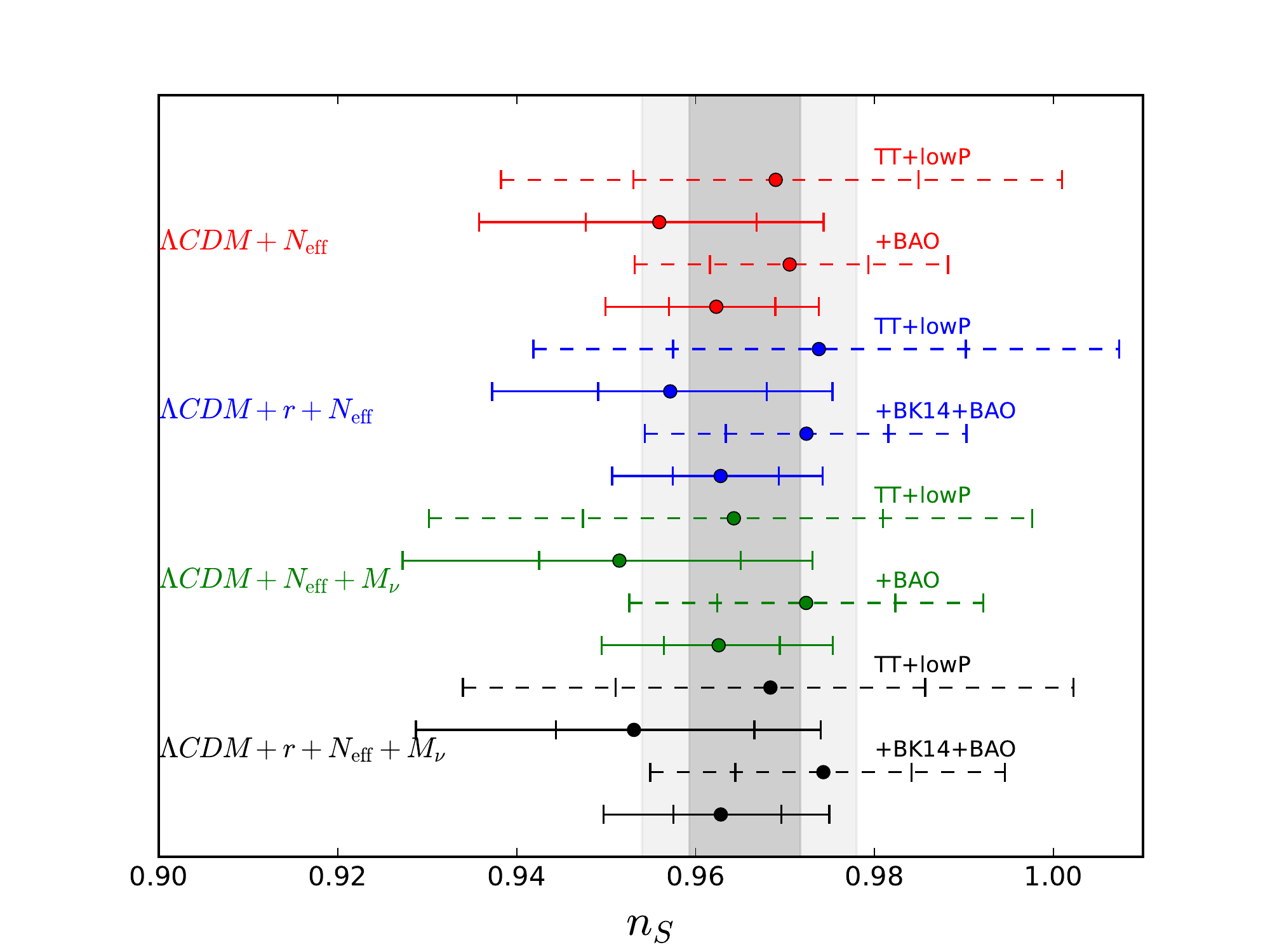}
\end{center}
\caption{Marginalized confidence intervals for the scalar spectral index $\ns$ for the indicated cosmological models and datasets. Solid bold lines are for the ``hard'' prior on $\Neff$ ($\Neff\leq3.046$), dashed lines are for ``broad'' marginalization ($0\leq\Neff\leq10$). The vertical bands are 68\% and 95\%~CL from Planck TT+lowP in the context of a $\lcdm$ model with one massive neutrino with mass $M_\nu=0.06\eV$ and $\Neff=3.046$.}\label{fig:1dns_nnu}
\end{figure}

The results are summarised in Tabs.~\ref{tab:ns_lcdmnnu} and also illustrated in Fig.~\ref{fig:1dns_nnu}. We report the 68\% CL around the mean. 
We firstly compare these results with those reported in Tab.~\ref{tab:ns_lcdm}, in order to highlight the effect of varying $\Neff$ instead of fixing it to the standard $\Neff=3.046$.
If we focus on the ``broad'' cases and compare them to the results obtained with the same combination of data and listed in Tab.~\ref{tab:ns_lcdm}, the net effect of allowing $\Neff$ to vary is to relax the bounds on $\ns$. The mean value of the probability distribution is shifted by about $0.2\sigma$ ($0.3\sigma$) towards higher values of $\ns$ for Planck TT+lowP (Planck TT + lowP+BAO) datasets in the $\lcdm+\Neff$ with respect to the baseline $\lcdm$ model. This effect is easily explained by recalling the strong degeneracy between $\Neff$ and $H_0$, which in turn is reflected on the bounds on $\ns$, as shown in Fig.~\ref{fig:tri_nnu}: when $\Neff$ is free to vary, more combinations of the parameters with respect to those accessible in the $\lcdm$ context become available which identify models in agreement with the data. Most importantly for comparison with inflation models, the uncertainty in $\ns$ becomes much larger, allowing for a wide range of $\ns$ (both larger and smaller than in standard $\Lambda$CDM). 

Moving to the ``hard'' cases, we would like to emphasise the fact that when $\Neff\leq3.046$, as expected in very low-reheating scenarios, the scalar spectral index can take quite small values. As an example, $\ns$ is shifted by $\sim0.8\sigma$ ($\sim0.7\sigma$) with respect to the value obtained in the baseline model for Planck TT+lowP (Planck TT+lowP+BAO). This is expected, given the degeneracy between $\ns$ and $\Neff$ discussed previously. In particular, decreasing $\Neff$ reduces the effect of the Silk damping (see e.g. \cite{2013PhRvD..87h3008H}), thus increasing the power at high multipoles, and this effect can be compensated by a smaller $\ns$.

Focusing now on the comparison between the two columns of Tab.~\ref{tab:ns_lcdmnnu}, we want to discuss the effect of varying the total neutrino mass in the context of a $\lcdm+\Neff$ scenario.
Looking at the results obtained in the ``broad'' scenario, when we marginalize over the total neutrino mass, the mean value of $\ns$ is lowered by a factor of $\sim0.3\sigma$ for Planck TT+lowP compared to the case of a fixed value of $M_\nu$. Interestingly, the sensitivity of the CMB data to $\ns$ (i.e. the size of the error bars) is the same for the cases of $\lcdm+ \Neff$ and
$\lcdm + \Neff+M_\nu$.
 In contrast, the constraints on the spectral index with BAO measurements included become weaker when $M_\nu$ is marginalized over. The reason for this behaviour resides in the effect that $M_\nu$ has on the $H_0$ and $\Omega_mh^2$ parameters, which are very well constrained by BAO data. Furthermore, a degeneracy between $\Neff$ and $M_\nu$ also appears in the BAO case, otherwise hidden when using CMB data only.

For the ``hard" case of $\Neff\leq3.046$, the preferred values
of $\ns$ are lower than for the standard case of $\Neff=3.046$ (the opposite direction of the shift
for the ``broad" case of arbitrary $\Neff$).
 In the case of $\lcdm+\Neff+M_\nu$ with Planck TT+lowP+BAO, $\ns$ is lowered by a factor of $\sim0.6\sigma$ with respect to standard $\lcdm$.
 
To summarise, the freedom induced by varying the number of relativistic degrees of freedom has a non-negligible impact on the bounds obtained on $\ns$. Indeed, the changes to the mean value of $\ns$
due to $\Neff$ are more relevant than those due to $M_\nu$.
Despite the fact that they lie on the exotic side of the inflationary possibilities, models with very low-reheating temperature can alter the thermalization of relativistic species, resulting in values of $\Neff$ lower than the standard 3.046 and  lower values of the scalar spectral index.

\begin{table}[h!]
\begin{tabular}{cc||c|c|}
&&$\Lambda\mathrm{CDM}+\Neff$	&$\Lambda\mathrm{CDM}+\Neff+M_\nu$\\
\hline
\multirow{2}{*}{TT+lowP} &broad &$0.969\pm 0.016$	& $0.964\pm 0.017$\\	
                  &hard	& $ 0.956^{+0.011}_{-0.0080}$		& $0.951^{+0.014}_{-0.0092}$\\
\hline
\multirow{2}{*}{+BAO} &broad  & $0.9705\pm 0.0089$	& $0.973\pm 0.010$\\	
                  &hard	& $ 0.9621^{+0.0067}_{-0.0053}$		& $0.9623^{+0.0068}_{-0.0057}$\\
\end{tabular}
\caption{68\%~CL probability intervals around the mean for the scalar spectral index $\ns$ for the indicated datasets and cosmological models. The rows labeled as ``broad'' refer to a full marginalization over $\Neff$ (\textit{i.e.} $\Neff$ free to vary within the range [0-10]), while the rows labeled as ``hard'' refer to models in which a hard prior on $\Neff$ has been adopted (\textit{i.e.} $\Neff\leq3.046$), in order to mimic low-reheating temperature inflationary scenarios.}\label{tab:ns_lcdmnnu}
\end{table}

\subsection{Relativistic degrees of freedom with a non-vanishing tensor component}
Here we extend the analysis of the previous section to include not only the effects of freeing up the values of $M_\nu$ and $\Neff$
but also allowing for a non-vanishing tensor component ($r\neq0$). The main results are reported in Tab.~\ref{tab:ns_lcdmrnnu}. As done for the $r=0$ case, we consider both the scenarios dummed as ``broad'' ($0\leq\Neff\leq10$) and ``hard'' ($\Neff\leq3.046$).

As already discussed, the inclusion of $r$ produces a slight enhancement of the mean value of $\ns$ with respect to the corresponding models in Tab.~\ref{tab:ns_lcdmnnu} with $r=0$. As a reference, $\ns$ increases by a factor of $\sim0.3\sigma$ moving from $\lcdm+\Neff$ to $\lcdm+r+\Neff$ for Planck TT+lowP in the ``broad'' case. The further marginalization over $M_\nu$, highlighted in the second column of Tab.~\ref{tab:ns_lcdmrnnu}, is once again responsible for a shift of $\ns$ towards lower values with respect to those obtained at fixed $M_\nu$ in the $\lcdm+r+\Neff$ context tested against CMB data (Planck TT+lowP alone or in combination with BK14). The inclusion of BAO helps constrain $H_0$, thus reducing the degeneracy with $\Neff$ and $\ns$. As a result, the constraints obtained with the Planck TT+lowP+BK14+BAO combination are the tightest among those reported in Tab.~\ref{tab:ns_lcdmrnnu}.

We confirm also in this case that the scenario represented by the ``hard'' marginalization prefers lower values of the scalar spectral index, for the very same reasons reported in the previous section.

\begin{table}[h!]
\begin{tabular}{cc||c|c|}
&&$\Lambda\mathrm{CDM}+r+\Neff$	&$\Lambda\mathrm{CDM}+r+\Neff+M_\nu$\\
\hline
\multirow{2}{*}{TT+lowP} &broad &$0.974\pm 0.016$	& $0.968^{+0.019}_{-0.017}$\\	
                  &hard	& $0.958^{+0.011}_{-0.0078}$		& $0.953^{+0.013}_{-0.0090}$\\
\hline
\multirow{2}{*}{+BK14} &broad &$0.972^{+0.015}_{-0.017}$	& $0.970\pm 0.016$\\	
                  &hard	& $0.956^{+0.011}_{-0.0079}$		& $0.954^{+0.012}_{-0.0080}$\\
\hline
\multirow{2}{*}{+BAO} &broad  & $0.9724\pm 0.0091$	& $0.974\pm 0.010$\\	
                  &hard	& $0.9629^{+0.0066}_{-0.0055}$		& $0.9628^{+0.0067}_{-0.0053}$\\
\end{tabular}
\caption{68\%~CL  probability intervals around the mean for the scalar spectral index $\ns$ for the indicated datasets and cosmological models. The rows labeled as ``broad'' refer to a full marginalization over $\Neff$ (\textit{i.e.} $\Neff$ free to vary within the range [0-10]), while the rows labeled as ``hard'' refer to models in which a hard prior on $\Neff$ has been adopted (\textit{i.e.} $\Neff\leq3.046$), in order to mimic low-reheating temperature  models.}\label{tab:ns_lcdmrnnu}
\end{table}

To conclude, the presence of non-vanishing tensor modes is again responsible for a shift in $\ns$ towards higher mean values with respect to the results obtained with $r=0$, also in the context of a free number of relativistic degrees of freedom. As a result, it might be relevant to take into account the effect of non-standard values of $\Neff$ when exploring the $\ns/r$ plane. Before moving to the following section, we would like to remind here some reference values of the scalar spectral index found in this analysis by employing current data. On one hand, we have $\ns=0.9656\pm0.0063$ in the baseline $\lcdm$ model, with $r=0$, $\Neff=3.046$ and $M_\nu=0.059\,\eV$ in the ``1+2'' parameterization (a similar value of $\ns=0.9655\pm0.0063$ has been found in the exact NH scenario). On the other hand, we have $\ns=0.9628^{+0.0067}_{-0.0053}$ in the most extended scenario of $\lcdm+\Neff+M_\nu+r$ with a ``hard'' prior on $\Neff$ tested against Planck TT+lowP+BK14+BAO. The very same scenario with a ``broad'' prior on $\Neff$ tested against the same combination of data provides instead the highest value of $\ns$: $\ns=0.974\pm0.010$ at 68\% CL.

\section{Forecasts}\label{sec:4cast}
In this section, we forecast the results expected from future CMB missions. We consider both the case of a future satellite mission, such as the proposed ESA satellite COrE~\cite{core}, and the case of a next-generation ground based observatory, such as the S4 telescope~\cite{Abazajian:2013oma,Abazajian:2013vfg,Abazajian:2016yjj}. We will again focus on the expected constraints on the scalar spectral index $\ns$ and discuss possible modifications to these constraints that might be induced by our imprecise knowledge of neutrinos properties, such as the unknown neutrino masses and hierarchy and/or a value of the relativistic degrees of freedom different from the standard one. Given the high sensitivity of future CMB measurements, these modifications might play a more important role than currently. 

When performing forecasts, one has to choose a fiducial model against which we compare various parameter studies.
In Tab.~\ref{tab:fiducial}, we report the values of the cosmological parameters defining our fiducial model. We choose a fiducial cosmology which assumes the NH scheme for the neutrino mass ordering, and a total neutrino mass $M_\nu=0.06\,\eV$, slightly higher than the minimal mass allowed by neutrino oscillation data in the NH scenario, such that all the three neutrino eigenstates are massive. We also assume a non-vanishing tensor component, corresponding to a value of the tensor-to-scalar ratio of $r=0.05$ at the pivot scale of $k=0.05\,\mathrm{Mpc^{-1}}$.

In addition to the fiducial cosmological model, one has also to simulate the data expected to be taken by the forecasted experiments, assuming the proposed experimental setup and expected noise properties.
In this work, the noise level is computed by adopting the experimental setup described in Table~\ref{tab:4cast_exp}, following the specifications from Ref.~\cite{Errard:2015cxa,rubmart}.

\begin{table}
\begin{tabular}{c|c}
$\Omega_bh^2$	&$0.02214$\\
$\Omega_ch^2$	&$0.1207$\\
$100\,\theta$	&$1.04075$\\
$\tau$	&$0.06$\\
$M_\nu\,[\eV]$	&$0.06$\\
$\Neff$	&$3.046$\\
$r$	&$0.05$\\
$\ns$	&$0.96$\\
$\ln[10^{10}A_s]$	&$3.053$\\
\end{tabular}\caption{Fiducial values for the cosmological parameters adopted when generating mock data. The total neutrino mass is distributed among the three massive eigenstates according to the normal hierarchy ordering.}\label{tab:fiducial}
\end{table}

\begin{table*}
\begin{tabular}{c||c|c}
	&COrE	&S4\\
\hline
\multirow{2}{*}{Frequency [GHz]}	&$(100,115,130,145$	&(90,150,220)\\
	                                                   &$160,175,195,220)$	&\\
\multirow{2}{*}{FWHM [arcmin]}	&$(8.4,7.3,6.46,5.79$	&$(5.0,3.0,2.0)$\\
	                                                   &$5.25,4.8,4.31,3.82)$	&\\
\multirow{2}{*}{T sensitivity \footnotemark[1] $\mathrm{[\mu K\,arcmin]}$}	&$(6.0,5.0,4.2,3.6$	&$(1.06,1.06,3.54)$\\
	                                                   &$3.8,3.8,3.85.8)$	&\\
Sky fraction	&$0.7$	&$0.5$\\
$\ell_\mathrm{min}-\ell_\mathrm{max}$	&$2-3000$	&$20-4000$\\
\end{tabular}
\footnotetext[1]{Temperature sensitivity. Polarization sensitivity is computed by rescaling the temperature sensitivity by a factor of $\sqrt{2}$.}\caption{Experimental setup adopted for the CMB forecasts analyzed in this work.}\label{tab:4cast_exp}
\end{table*}

We test our fiducial model against both COrE and S4 (complemented with a Gaussian prior of $\tau=0.06\pm0.01$ on the optical depth), assuming three different cosmological models: i) $\lcdm+r$, with $M_\nu$ fixed to the fiducial value ($M_\nu=0.06\,\eV$) and $\Neff$ fixed to the standard expected value ($\Neff=3.046$); ii) $\lcdm+r+M_\nu$, same as i) but with $M_\nu$ free to vary; iii) $\lcdm+r+\Neff$, same as i), but with $\Neff$ free to vary subjected to the hard prior of $\Neff\leq3.046$. In all the cases the hierarchy has been chosen to be normal.

We report our results in Tab.~\ref{tab:ns_4cast}. The main message of Tab.~\ref{tab:ns_4cast} is that we still recover a lower value for the scalar spectral index once the total neutrino mass and/or the number of relativistic degrees of freedom (for the hard prior of $\Neff\leq3.046$) are marginalized over. This is particularly evident for the COrE-like case. Indeed, if we take as a reference the value recovered in the $\lcdm+r$ scenario (first line of Tab.~\ref{tab:ns_4cast}), $\ns=0.9601\pm0.0014$, we find a $\sim0.4\sigma$ ($\sim0.8\sigma$) shift towards lower values when moving from the $\lcdm+r$ to the $\lcdm+r+M_\nu$ ($\lcdm+r+\Neff$) model.

We can also ask ourselves if a wrong assumption about the exact neutrino mass splitting might induce a bias in the recovered value of $\ns$ and/or degrade the uncertainty on $\ns$. In particular, we would like to address the question whether we can still make use of the approximate parameterization rather than follow an exact scheme, as usually done in cosmological analysis. Therefore, we perform the following test: we assume a $\lcdm+r$ scenario, with $M_\nu$ fixed to the fiducial value \textit{but} distributed among the eigenstates according to the ``1+2'' parameterization (remember that the fiducial choice for the neutrino mass splitting was the NH scenario).
%
We find the following two values for the spectral index (which are not reported in Tab.\ref{tab:ns_4cast}):  $\ns=0.9598\pm 0.0014$ for COrE and $\ns=0.9600\pm 0.0019$ for S4, both at 68\% CL assuming ``1+2'' parameterization instead of NH, to be compared, along with the equivalent results reported in Tab.\ref{tab:ns_4cast} for the NH parameterization ($\ns=0.9601\pm0.0014$ for COrE and $\ns=0.9599\pm0.0019$ for S4), with the fiducial value assumed for the scalar spectral index, $\ns=0.96$. These results highlight that there is an almost negligible impact of the underlying mass-splitting parameterization on the spectral index, when dealing with future CMB data. Once again, we have found that the exact choice of the neutrino parameterization does not have significant effect on the stability of the constraints on $\ns$. Nevertheless, we would like to notice that we recover the same pattern already present in Tab.~\ref{tab:ns_lcdmr}, with the ``1+2'' parameterization favouring overall a (slightly) lower value of $\ns$.

\begin{table}[h!]
\begin{tabular}{c||c|c|}
&COrE	&S4\\
\hline
$\lcdm+r$ 	&$0.9601\pm 0.0014$	&$0.9599\pm 0.0019$\\
$\lcdm+r+M_\nu$	&$0.9593\pm 0.0016$	&$0.9595\pm 0.0020$\\
$\lcdm+r+\Neff^\mathrm{h}$ &$0.9580^{+0.0024}_{-0.0017}$	&$0.9580^{+0.0027}_{-0.0023}$\\
\end{tabular}
\caption{68\%~CL probability intervals around the mean for the scalar spectral index $\ns$ for the indicated CMB datasets (the satellite mission COrE and the ground based telescope S4) and cosmological models. S4 is complemented by a Gaussian prior on the reionization optical depth $\tau=0.06\pm0.01$. The superscript ``h'' in $\Neff$ indicates that we are reporting results for the ``hard'' marginalization over $\Neff$ ($\Neff\leq3.046$).}\label{tab:ns_4cast}
\end{table}

To summarize this section, the largest bias on $\ns$ is induced by relaxing the assumptions about the number of relativistic degrees of freedom. Nevertheless, even assuming that we were able to determine $\Neff$ with infinite precision, thus completely erasing its impact on $\ns$, an imperfect knowledge of the total neutrino mass would still induce a $\sim0.4\sigma$ bias in the recovered value of $\ns$, as highlighted by the results obtained by marginalizing over $M_\nu$ and reported in Tab.\ref{tab:ns_4cast}.

\section{Implications for Specific Inflationary Models}\label{sec:inflation}
The results reported above may have implications when discussing the ability of cosmological data to discriminate among different inflationary models, particularly for future experiments.  The forecasted sensitivity on the inflationary parameters and the consequent significance with which some inflationary models are expected to be ruled out should be carefully assessed, including all possible uncertainties.

Here, we consider the implications of our findings for a few selected, theoretically well-motivated, inflationary models, which could be prematurely discarded if biases due to uncertainties in the neutrino sector are not carefully taken into account when estimating inflationary parameters from cosmological datasets. The panels from Fig.~\ref{fig:2d_Nat} to Fig.~\ref{fig:2d_4c_Higgs} depict the two-dimensional contours at 68\% and 95\% CL in the $\ns/r$ plane for a selection of models and datasets. The datasets have been discussed in Sec.~\ref{sec:method}, whereas the selection of theoretical models will be briefly outlined in what follows.

\subsection{Natural inflation (NI)}

In order to satisfy constraints on sufficient inflation and anisotropy in the CMB, the potential for the inflaton must be very flat, in the sense that the ratio of the height to the (width)$^4$ of the potential has to be of order ${\cal O}(\ll 1)$, which in turn implies that the inflaton effective quartic self-coupling $\lambda$ must be comparably small. From a particle physics standpoint, a theoretically desirable situation is that where the smallness of $\lambda$, and hence the flatness of the potential, is protected by a symmetry, and hence natural in the sense of 't Hooft~\cite{hooft}. In natural inflation \cite{Freese:1990rb,Adams:1992bn}, the role of the inflaton is played by a Pseudo-Nambu-Goldstone-Boson (PNGB) $\phi$, such as the axion (although not the QCD axion). The key ingredient in keeping the potential flat is
a shift symmetry.  As long as the shift symmetry is exact, the inflaton
cannot roll and drive inflation, and hence there must be additional
explicit symmetry breaking.  Then these particles become pseudo-Nambu
Goldstone bosons (PNGBs), with nearly flat potentials, exactly as
required by inflation. 

\subsubsection{Cosine Natural Inflation}
In the original Cosine Natural Inflation model, modelled after the QCD axion,
 the PNGB potential resulting from explicit breaking of a shift symmetry is  of
the form
\begin{eqnarray}
V(\phi) = \Lambda^4 [1 \pm \cos(N\phi/f)] \, .
\label{potential}
\end{eqnarray}
We will take the positive sign in Eq.(\ref{potential}) (this choice has
no effect on our results) and take $N = 1$, so the potential, of
height $2 \Lambda^4$, has a unique minimum at $\phi = \pi f$ (the
periodicity of $\phi$ is $2 \pi f$).
For appropriately chosen values of the mass scales, e.g. $f \geq \mpl$
and $\Lambda \sim \mgut \sim 10^{16}$ GeV, the PNGB field $\phi$ can
drive inflation.  Then the inflaton mass
$m_\phi = \Lambda^2/f \sim 10^{11}$-$10^{13}$~GeV.    For $f \gg \mpl$, the inflaton
becomes independent of the scale $f$ and is $m_\phi \sim 10^{13}$ GeV.
For $f \gg \mpl$, the predictions of natural inflation tend to those of the minimally coupled quadratic chaotic inflation model, i.e. with a potential $V \propto \phi^2$.

In this paper we take $\mpl = 1.22 \times 10^{19}$ GeV.  Our studies of natural inflation in light of data
extend upon previous analyses of NI \cite{Freese:2004un} and \cite{Savage:2006tr} 
that was based upon WMAP's first year data \cite{Spergel:2003cb} and third year data and \cite{Freese:2014nla} based on Planck data.
Even earlier analyses \cite{Adams:1992bn,Moroi:2000jr}  placed
observational constraints on this model using COBE data
\cite{Smoot:1992td}.  Other papers have studied inflation models (including NI)  in light of the WMAP1 and WMAP3 data
\cite{Alabidi:2006qa,Alabidi:2005qi} and in light of Planck data \cite{Ade:2015lrj,Planck:2013jfk,Tsujikawa:2013ila}.
Huang et al. \cite{Huang:2015cke} found that, in the context of $\lcdm+r$ model and two extended models ($\lcdm+r+n_\mathrm{run}$ and $\lcdm+r+n_\mathrm{run}+n_\mathrm{run,run}$, where $n_\mathrm{run}=d\ns/d\ln k$ is the running of $\ns$ and $n_\mathrm{run,run}=d^2\ns/d\ln k^2$ is the running of the running) Cosine Natural Inflation is in tension at the two sigma level
with a combination of  Planck, BICEP2/Keck Array \cite{Array:2015xqh}, and BAO data.  

We reinvestigate this here by allowing for more extended cosmological models which take into account possible deviations from the standard value of relativistic degrees of freedom $\Neff$, and the uncertainties related to the value of the total neutrino mass as well as the distribution of the total mass among the neutrino eigenstates. We find basic agreement with the conclusion of Ref.~\cite{Huang:2015cke} for the case of standard $\Neff$.

\begin{figure*}
\begin{tabular}{cc}
\includegraphics[width=0.5\textwidth]{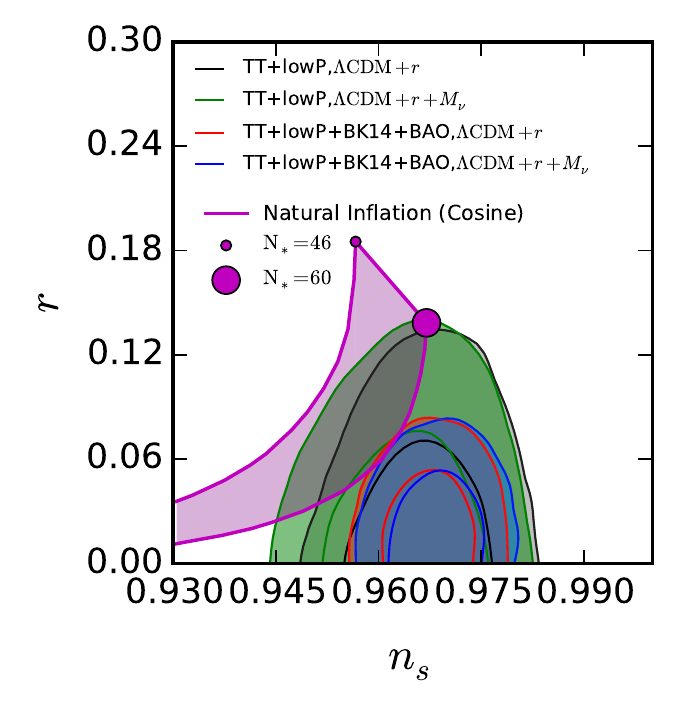}&\includegraphics[width=0.5\textwidth]{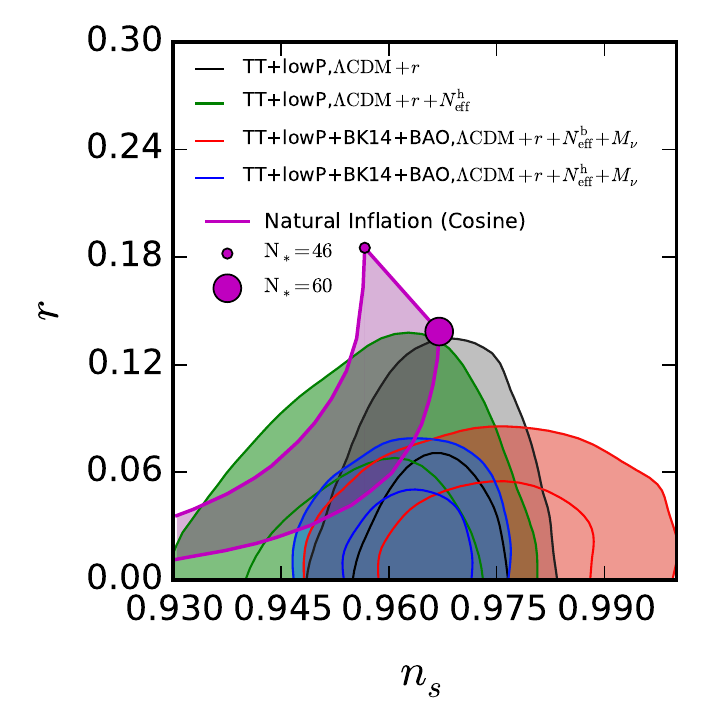}\\
\end{tabular}
\caption{Two-dimensional probability contours at 68\% and 95\% CL in the $\ns/r$ plane for the indicated datasets and models. The purple region shows predictions for Cosine Natural Inflation models for $46\leq N_*\leq60$, where $N_*$ is the number of e-folds up to the end of inflation at which present modes of $k=0.002\,\mathrm{Mpc}^{-1}$ have been generated. \textit{Left panel:} focus on the impact of neutrino hierarchy and total mass in the $\ns/r$ plane. Contours are drawn assuming the ``NH'' parameterization. \textit{Right panel:} focus on the impact of varying the number of relativistic degrees of freedom in the $\ns-r$ plane. The superscripts ``h'' and ``b'' stand for the hard ($\Neff\leq3.046$) and broad ($0\leq\Neff\leq10$) prior imposed to $\Neff$. Contours are drawn assuming either the ``1+2'' parameterization when $M_\nu$ is fixed or the ``3deg'' parameterization when $M_\nu$ is marginalized over.}\label{fig:2d_Nat}
\end{figure*}

\begin{figure*}
\begin{tabular}{cc}
\includegraphics[width=0.5\textwidth]{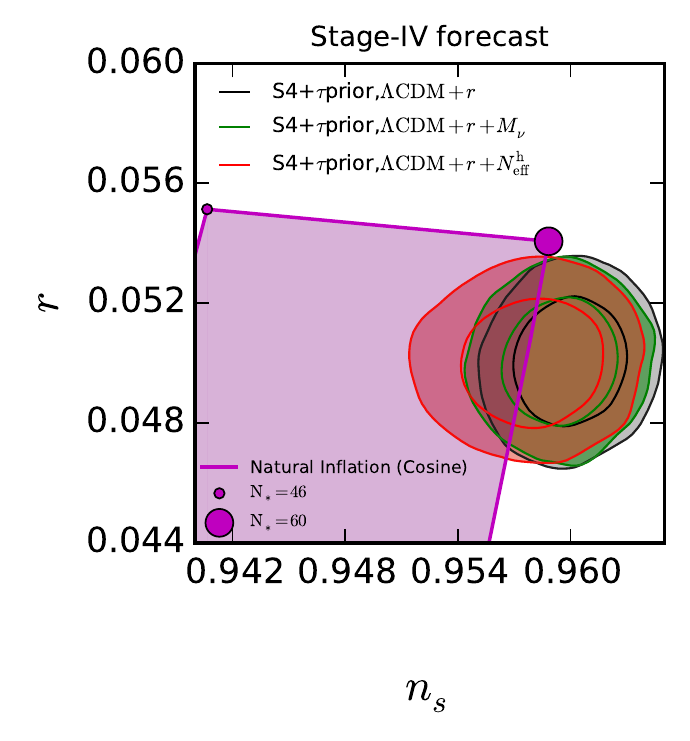}&\includegraphics[width=0.47\textwidth]{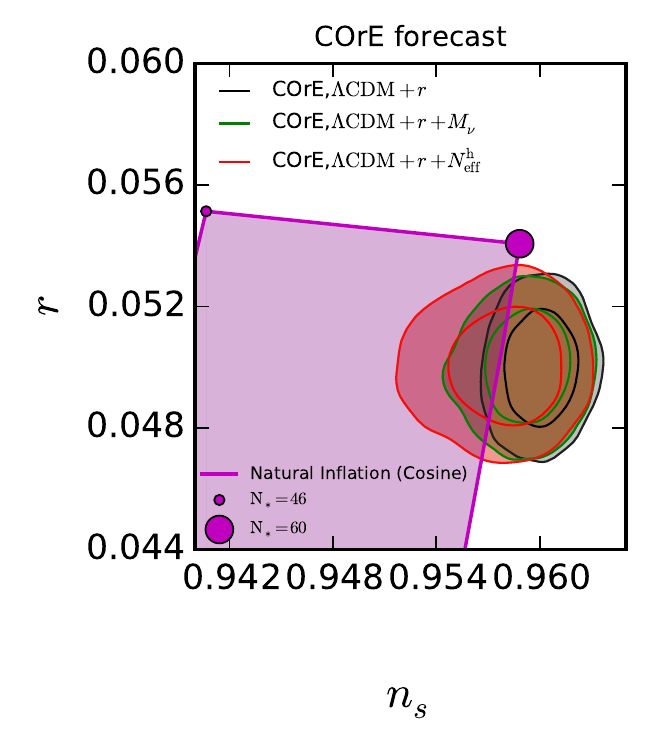}\\
\end{tabular}
\caption{Two-dimensional probability contours at 68\% and 95\% CL in the $\ns/r$ plane for the indicated datasets and models. Contours are drawn assuming the ``NH'' parameterization and a cosmological model with $\ns=0.96$ and $r=0.05$. The purple region shows predictions for Natural Inflation model for $46\leq N_*\leq60$, where $N_*$ is the number of e-folds up to the end of inflation at which present modes of $k=0.002\,\mathrm{Mpc}^{-1}$ have been generated.  \textit{Left panel:} focus on forecasted results from the combination of the next-generation ground-based CMB experiment Stage-IV and a Gaussian prior on the optical depth $\tau=0.06\pm0.01$. \textit{Right panel:} focus on forecasted results from the  next-generation satellite CMB experiment COrE.}\label{fig:2d_4c_Nat}
\end{figure*}

The predictions for Cosine Natural inflation in the $\ns/r$ plane are plotted as a purple band in both panels of Fig. ~\ref{fig:2d_Nat}, together with the two-dimensional 68\% and 95\% CL contours for the indicated datasets and models. The left panel shows extensions of the $\lcdm+r$ model while taking $\Neff$ fixed to $\Neff=3.046$ and assuming that the total neutrino mass is distributed according to the ``NH'' scenario. For Planck measurements only, for the case of $\lcdm+r$, the grey region shows that the Cosine NI model is in tension at 2 sigma.  The marginalization over $M_\nu$ with $\Neff=3.046$ shifts the mean value of $\ns$ to lower values so that the agreement is enhanced.    The inclusion of BK14 and BAO measurements, however, shifts the mean value back to higher values, reinstating the $2\sigma$ tension. A similar plot may be found
 in Planck 2013 paper \cite{Planck:2013jfk}, though there the assumption was made of
 three degenerate neutrino species (for $\lcdm+r+M_\nu$) and ``1+2'' species (for $\lcdm+r$), while here we report contours by assuming the exact NH parameterization.


The right panel shows results obtained by allowing a free value of $\Neff$. In low-reheating scenarios with $\Neff$ below the standard value (indicated by the ``h'' superscript), Cosine NI is within the 68\% contours; however, this is a somewhat exotic case. Notice also that 
the model just reaches $2\sigma$ agreement again for the case where the $\lcdm+r+\Neff+M_\nu$ model is tested against the full combination of CMB and BAO data (red contours).

Finally, with regard to forecasts for future surveys (see Fig.~\ref{fig:2d_4c_Nat}), we again emphasise the effect of neutrino properties on the remaining regions in the $\ns/r$ plane.
 
\subsubsection{Generalizations of Original Natural Inflation}

Subsequent to the original Cosine variant of Natural Inflation, 
many types of candidates have subsequently been explored for natural
inflation.  For example, as discussed in the next section, Kinney and Mahanthappa considered NI
potentials generated by radiative corrections in models with explicitly
broken Abelian \cite{Kinney:1995xv} and non-abelian \cite{Kinney:1995cc}
symmetries. 
Cosine NI requires the width of the
potential to be trans-Planckian.  Such a scenario is difficult to accommodate in
string theory.  Thus many authors have proposed other variants of NI, taking advantage of the shift
symmetry offered by "axions," and looking for extensions
of the original cosine potential that accommodate smaller values of $f$ \footnote{Kim, Nilles \& Peloso \cite{Kim:2004rp} as well as the idea of N-flation
\cite{Dimopoulos:2005ac,Easther:2005zr,Grimm:2007hs} generalized the original NI model to include
two or more axions, and showed that an \textit{effective} potential of
$f \gg \mpl$ can be generated from multiple axions, each with
sub-Planckian scales. An interesting variant is modulated natural inflation 
\cite{Kappl:2015esy}.

Ref.~\cite{Kawasaki:2000yn} used shift symmetries
in K\"ahler potentials to obtain a flat potential and drive natural
chaotic inflation in supergravity.  Additionally,
\cite{ArkaniHamed:2003wu,ArkaniHamed:2003mz} examined natural
inflation in the context of extra dimensions and \cite{Kaplan:2003aj}
used PNGBs from little Higgs models to drive hybrid inflation.  Also,
\cite{Firouzjahi:2003zy,Hsu:2004hi} use the natural inflation idea of
PNGBs in the context of braneworld scenarios to drive inflation.
Freese \cite{Freese:1994fp} suggested using a PNGB as the rolling
field in double field inflation \cite{Adams:1990ds} (in which the
inflaton is a tunnelling field whose nucleation rate is controlled by
its coupling to a rolling field).  Models have been proposed with enhanced friction occurring during axion inflation \cite{Anber:2009ua,Germani:2010hd}.  Ref. \cite{Kaloper:2008fb,Kaloper:2011jz} found
 a quadratic potential  in theories where an "axion" field mixes with a 4-form.
Ref. \cite{Germani:2010hd,Germani:2011ua} used coupling of 
the inflaton kinetic term to the Einstein tensor to allow NI with $f << \mpl$ by enhancing the gravitational friction
 acting on the inflaton during inflation.   Ref. \cite{Czerny:2014wza,Czerny:2014xja} suggested a
 "multi-natural" inflation model in which the single-field inflaton potential consists of two or more 
 sinusoidal potentials  with a possible non-zero relative phase
 (such as may arise if a complex scalar field  is coupled to two sets of quark and anti- quark fields).
 We will focus in this paper on 
single field implementations of NI.}.
Axion monodromy \cite{Silverstein:2008sg} is a shift-symmetric string-motivated extension of Natural Inflation which evades the super-Planck scale width of the cosine potential by analytic continuation on a compact manifold, resulting in an effective field range larger than $\mpl$.  A resulting potential is $V \propto \phi^{2/3}$  or in Linear Axion Monodromy \cite{McAllister:2008hb}, $V \propto \phi$, see also
\cite{Flauger:2009ab,Berg:2009tg}. Versions of axion monodromy with additional couplings to heavy degrees of freedom can produce larger tensor amplitudes \cite{Dong:2010in}.

\subsubsection{KM  Natural Inflation --- Quartic Hilltop}

In 1995, Kinney and Mahanthappa proposed a realization of 
low-scale inflationary scenarios wherein the inflaton potential is generated by radiative corrections in an explicitly broken SO(3) gauge symmetry. Within the model, which we refer to as KM natural inflation, the inflaton is a pseudo-Goldstone mode whose dynamics are governed by the following potential
\begin{equation}
V = V_0  \left\lbrace \sin^4\left(\frac{\phi}{\mu}\right) \log\left[g^2 \sin^2\left(\frac{\phi}{\mu}\right)\right] - \log\left[g^2\right]\right\rbrace.
\end{equation}
In the low-scale limit $\mu \ll \mpl$, potentials of this type reduce to the quartic hilltop model at leading order in a Taylor expansion,
\begin{equation}
V \simeq V_0 - \lambda \left(\frac{\phi}{\mu}\right)^4 + \cdots
\end{equation}
Extension of the model to arbitrary values of $\mu$ (sub- or super-Planckian) are possible
and compatible with data.

\begin{figure*}
\begin{tabular}{cc}
\includegraphics[width=0.5\textwidth]{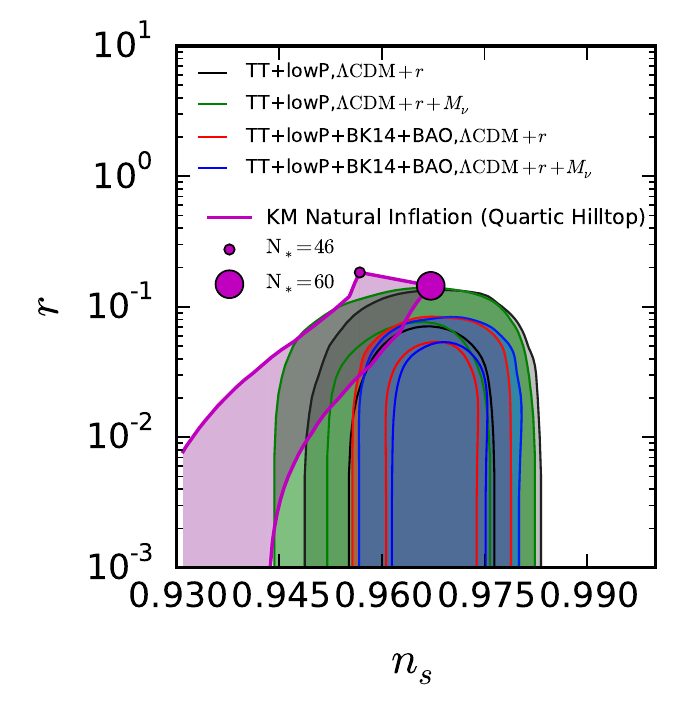}&\includegraphics[width=0.5\textwidth]{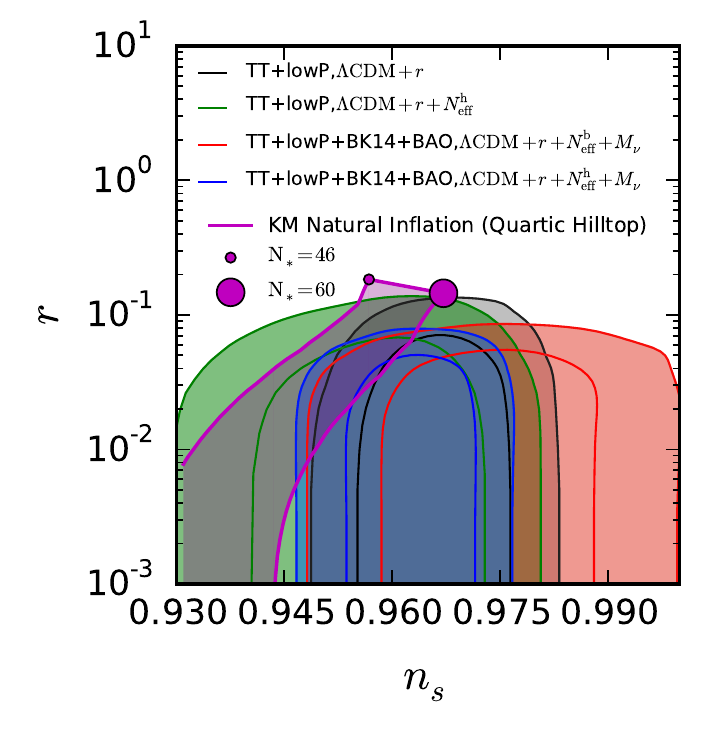}\\
\end{tabular}
\caption{Two-dimensional probability contours at 68\% and 95\% CL in the $\ns/r$ plane for the indicated datasets and models. The purple region shows predictions for the Kinney-Mahanthappa (KM) model of Natural Inflation (or Quartic Hilltop) for $46\leq N_*\leq60$, where $N_*$ is the number of e-folds up to the end of inflation at which present modes of $k=0.002\,\mathrm{Mpc}^{-1}$ have been generated. \textit{Left panel:} focus on the impact of neutrino hierarchy and total mass in the $\ns-r$ plane. Contours are drawn assuming the ``NH'' parameterization. \textit{Right panel:} focus on the impact of varying the number of relativistic degrees of freedom in the $\ns/r$ plane. The superscripts ``h'' and ``b'' stand for the hard ($\Neff\leq3.046$) and broad ($0\leq\Neff\leq10$) prior imposed to $\Neff$. Contours are drawn assuming either the ``1+2'' parameterization when $M_\nu$ is fixed, or the ``3deg'' parameterization when $M_\nu$ is marginalized over.}\label{fig:2d_KM}
\end{figure*}

In Fig.~\ref{fig:2d_KM}, we show the predictions from KM Natural Inflation as a purple band and their agreement with a selection of models/datasets discussed in this work. As for Fig.~\ref{fig:2d_Nat}, the left panel shows extensions to $\lcdm+r$ with fixed $\Neff$, while the right panel also reports results for a free $\Neff$. Also in this case, there is agreement at $2\sigma$ level between KM predictions and the allowed contours in the $\ns/r$ plane in the context of $\lcdm+r$ and $\lcdm+r+M_\nu$ when the full combination of CMB and BAO data is employed (see red and blue regions in the left panel of Fig.~\ref{fig:2d_KM}). Better agreement is recovered when also $\Neff$ is marginalized over (red contours in the right panel) and in the (more exotic) context of low-reheating scenarios (green and blue regions in the right panel).

\begin{figure*}
\begin{tabular}{cc}
\includegraphics[width=0.5\textwidth]{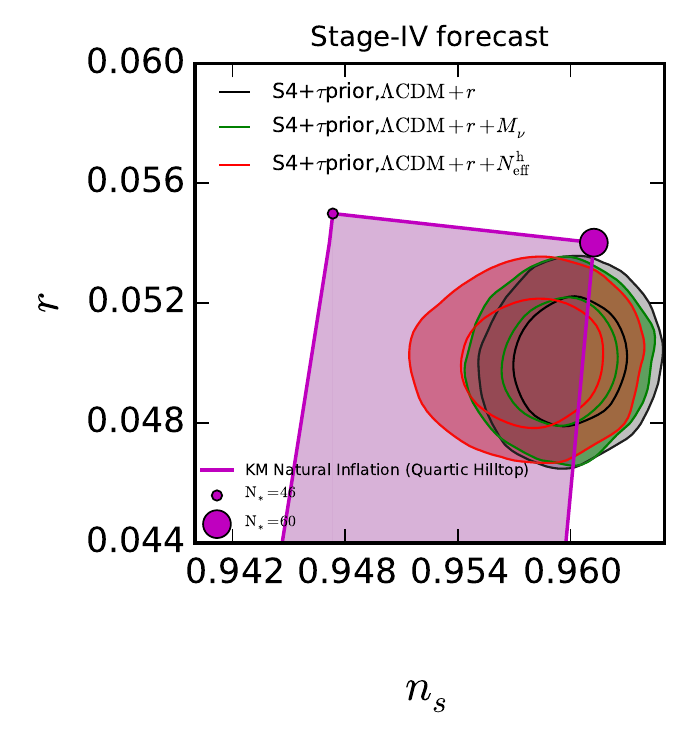}&\includegraphics[width=0.47\textwidth]{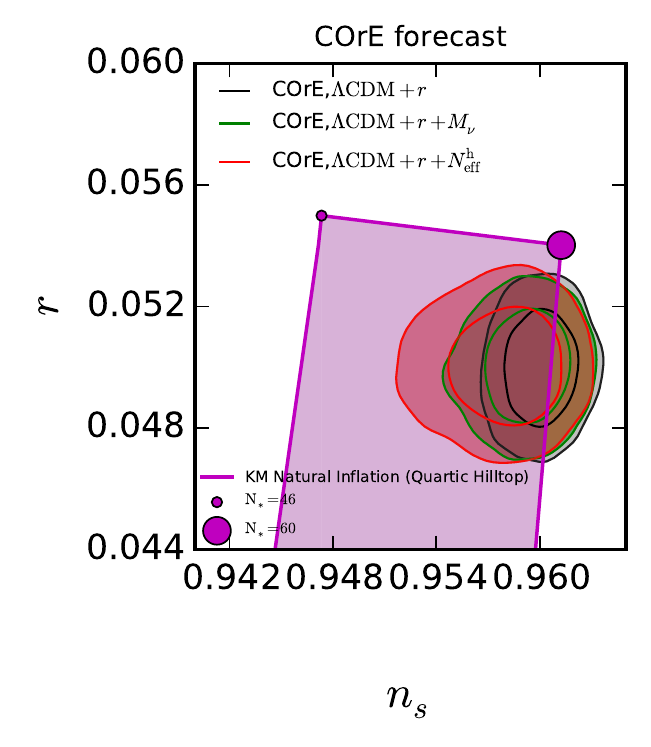}\\
\end{tabular}
\caption{Two-dimensional probability contours at 68\% and 95\% CL in the $\ns/r$ plane for the indicated datasets and models. Contours are drawn assuming the ``NH'' parameterization and a cosmological model with $\ns=0.96$ and $r=0.05$. The purple region shows predictions for Kinney-Mahanthappa (KM) model of Natural Inflation (or Quartic Hilltop) for $46\leq N_*\leq60$, where $N_*$ is the number of e-folds up to the end of inflation at which present modes of $k=0.002\,\mathrm{Mpc}^{-1}$ have been generated.  \textit{Left panel:} focus on forecasted results from the combination of the next-generation ground-based CMB experiment Stage-IV and a Gaussian prior on the optical dept $\tau=0.06\pm0.01$. \textit{Right panel:} focus on forecasted results from the  next-generation satellite CMB experiment COrE.}\label{fig:2d_4c_KM}
\end{figure*}

Similar conclusions can be drawn with regard to the forecasts reported in Fig.~\ref{fig:2d_4c_KM}: the marginalization over $M_\nu$ and $\Neff$ extends the allowed region of the parameter space towards lower values of $\ns$, recovering a better agreement with KM predictions.

\subsection{Higgs-like models}

Here we study the potential of a Higgs-like particle at the GUT scale~\cite{Kaplan:2003aj} \footnote{For the sake of clarity, we remark that this model is distinct from the Higgs inflation model formulated in~\cite{Bezrukov:2007ep}, where it is the actual Higgs boson (and not a Higgs-like particle) which drives inflation, through a non-minimal coupling to the Ricci scalar (see also~\cite{Escudero:2015wba}).}:

\begin{equation}
V\left(\phi\right) = V_0 \left[1 - \left(\frac{\phi}{\mu}\right)^2\right]^2
\end{equation}

\begin{figure*}
\begin{tabular}{cc}
\includegraphics[width=0.5\textwidth]{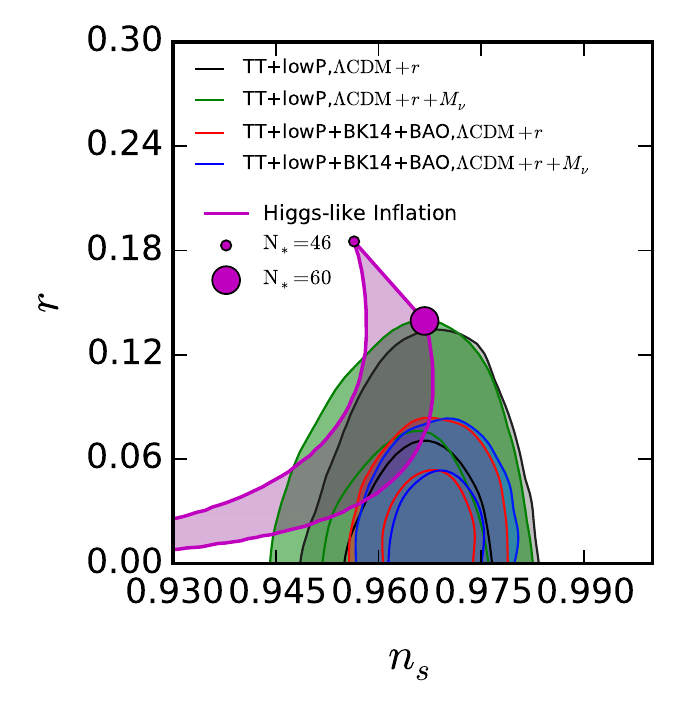}&\includegraphics[width=0.5\textwidth]{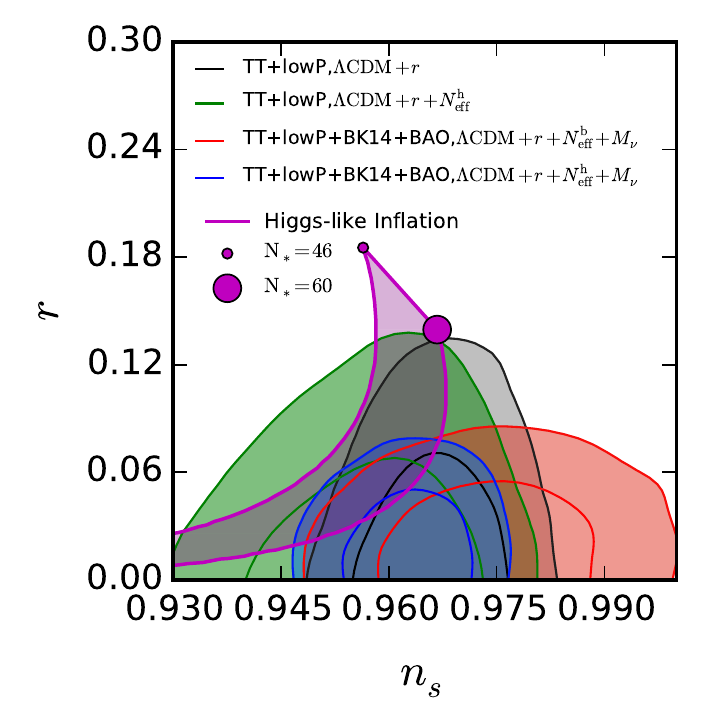}\\
\end{tabular}
\caption{Two-dimensional probability contours at 68\% and 95\% CL in the $\ns/r$ plane for the indicated datasets and models. The purple region shows predictions for Higgs-like models of inflation for $46\leq N_*\leq60$, where $N_*$ is the number of e-folds up to the end of inflation at which present modes of $k=0.002\,\mathrm{Mpc}^{-1}$ have been generated. \textit{Left panel:} focus on the impact of neutrino hierarchy and total mass in the $\ns/r$ plane. contours are drawn assuming the ``NH'' parameterization. \textit{Right panel:} focus on the impact of varying the number of relativistic degrees of freedom in the $\ns-r$ plane. The superscripts ``h'' and ``b'' stand for the hard ($\Neff\leq3.046$) and broad ($0\leq\Neff\leq10$) prior imposed to $\Neff$. The contours are drawn assuming  either the ``1+2'' parameterization when $M_\nu$ is fixed or the ``3deg'' parameterization when $M_\nu$ is marginalized over.}\label{fig:2d_Higgs}
\end{figure*}

The comparison between predictions of the model and regions in the $\ns/r$ plane allowed by current data is reported in Fig.~\ref{fig:2d_Higgs}, where theoretical predictions are shown as a purple band. Better agreement is obtained in this case. The full combination of CMB and BAO data still limits the agreement to a reduced region of the parameter space (see red and blue contours in the two panels of Fig.~\ref{fig:2d_Higgs}).

\begin{figure*}
\begin{tabular}{cc}
\includegraphics[width=0.5\textwidth]{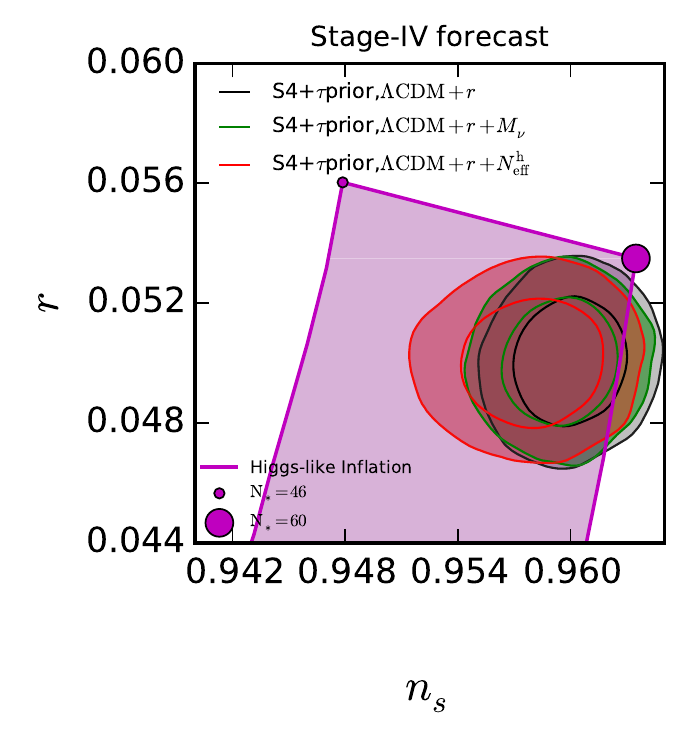}&\includegraphics[width=0.47\textwidth]{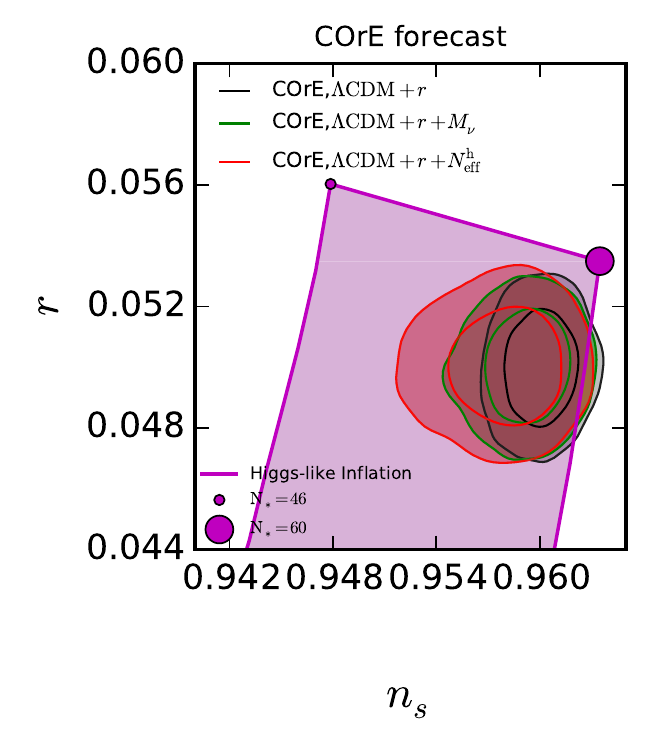}\\
\end{tabular}
\caption{Two-dimensional probability contours at 68\% and 95\% CL in the $\ns/r$ plane for the indicated datasets and models. Contours are drawn assuming the ``NH'' parameterization and a cosmological model with $\ns=0.96$ and $r=0.05$. The purple region shows predictions for Higgs-like models of inflation for $46\leq N_*\leq60$, where $N_*$ is the number of e-folds up to the end of inflation at which present modes of $k=0.002\,\mathrm{Mpc}^{-1}$ have been generated.  \textit{Left panel:} focus on forecasted results from the combination of the next-generation ground-based CMB experiment Stage-IV and a Gaussian prior on the optical dept $\tau=0.06\pm0.01$. \textit{Right panel:} focus on forecasted results from the  next-generation satellite CMB experiment COrE.}\label{fig:2d_4c_Higgs}
\end{figure*}

Concerning forecasts, Fig.~\ref{fig:2d_4c_Higgs} shows once again that the marginalization over the neutrino properties analysed in this work (namely, the total neutrino mass and the number of relativistic degrees of freedom) enlarges the allowed probability region of the parameter space in the $\ns/r$ plane, which is also in agreement with the predictions from the Higgs-like model of inflation.

\section{Conclusions}\label{sec:conclusion}


We have investigated here the robustness of the constraints on the scalar spectral index $\ns$ under various assumptions about neutrino properties, by employing a combination of CMB and BAO data in the context of a $\lcdm$ model and other possible extensions. In particular, we have considered the impact of marginalizing over the total neutrino mass $M_\nu$, under different choices for the neutrino mass splittings. We have compared the results arising from assuming either an approximate neutrino mass-splitting (one massive eigenstate carrying the total mass plus two massless species when the total mass is fixed to the minimal-mass case, $M_\nu=0.059\,\eV$, and three degenerate massive neutrinos otherwise) as usually done in literature, or the exact mass-splitting (normal and inverted neutrino mass hierarchies). We have found that the assumptions about the mass splittings play a negligible role in the context of current cosmological measurements. However, when values of the neutrino mass different from the minimal mass of $M_\nu\sim0.059\eV$ are taken into account, the spectral index is slightly shifted \footnote{In this context, the choice of the hierarchy, either exact or approximate, does play a role, but mainly due to the different prior on $M_\nu$ adopted in the two cases ($M_\nu>M_{\nu,\mathrm{min}}$, with $M_{\nu,\mathrm{min}}\neq0$ in the exact parameterization but $M_{\nu,\mathrm{min}}=0$ for the approximate parametrization).}. This is due to a mild inverse degeneracy between the two parameters, induced by the strong degeneracy between $M_\nu$, the Hubble constant $H_0$ and the matter density $\Omega_mh^2$. These degeneracies can be strongly alleviated by the addition of BAO measurements. 

We have also tested the effect of considering a free number of relativistic degrees of freedom $\Neff$ and found that the scalar spectral index is considerably lowered when $\Neff\leq3.046$, as expected in the context of low-reheating temperature scenarios ($T_\mathrm{RH}\sim\mathcal{O}(\mathrm{MeV})$, subjected to $T_\mathrm{RH}>5\,\mathrm{MeV}$ in order to satisfy BBN bounds~\cite{deSalas:2015glj}). This shift in $\ns$ is mostly driven by the strong degeneracy between $\Neff$, $\ns$ and $H_0$. This preference is mildly alleviated by the inclusion of BAO data, which are able to exclude low values of $H_0$ \footnote{We have not included priors from direct measurements of $H_0$ on purpose, given the highly discussed tension with the CMB determination of the same quantity (see \cite{Riess:2016jrr,DiValentino:2016hlg,Qing-Guo:2016ykt,Bernal:2016gxb} and references therein). However, we can qualitatively comment on the possible effect of including priors from the local measurements of the Hubble constant: the aforementioned priors, preferring preferring values of the Hubble constant $H_0$ higher than those predicted by CMB and BAO data, would go in the direction of favoring, on average, higher values of $\ns$, given the direct degeneracy between the two parameters. In addition, it would also lower the upper bound on $M_\nu$ and increase the value of $\Neff$, favouring again higher values of $\ns$.}.

We have also allowed for a non-vanishing tensor component in the analyses (namely, we have investigated $\lcdm+r$ models and discussed extensions), finding that the inclusion of non-vanishing tensor modes is responsible for slightly increasing $\ns$ with respect to the corresponding bounds with $r=0$. 

The dependence of the constraints of $\ns$ on neutrino properties, especially in extended $\lcdm+r$ scenarios, is crucial for assessing the significance at which inflationary models can be excluded with cosmological data. As an example, $\sim 1\sigma$ agreement between the predictions from the cosine natural inflation paradigm with Planck TT+lowP+BK14+BAO data can be recovered in exotic reheating scenarios with $\lcdm+r+M_\nu+\Neff$. In addition, it will not be possible to exclude the model with future CMB data alone, if one performs forecasts of future CMB missions such as COrE and Stage-IV by assuming a fiducial model with $\ns=0.96$ and $r=0.05$. Similar considerations hold for the KM variant of Natural Inflation, as well as for Higgs-like models of inflation.  

A precise determination of both the mean value and the error budget associated to a determination of $\ns$ by including all the possible sources of uncertainties is therefore a mandatory analysis. The very same forecasts of future CMB missions discussed above show that constraints on $\ns$ can be altered by more than $\sim0.4\sigma$ if uncertainties related to our incomplete knowledge of the neutrino properties (\textit{i.e.} the precise value of the total mass and exact number of relativistic degrees of freedom) are not taken into account properly. This is crucial for upcoming experiments aiming at the discovery of the inflationary paradigm, given the claimed precision at which they would be able to constrain the inflationary sector.

\begin{acknowledgments}
This work is based on observations obtained with Planck (http://www.esa.int/Planck), an ESA science mission with instruments and contributions directly funded by ESA Member States, NASA, and Canada. We also acknowledge use of the Planck Legacy Archive.
K. F. acknowledges support from DoE grant DE-SC0007859 at the University of Michigan as well as support from the 
Michigan Center for Theoretical Physics.
M.G., S.V. and K.F. acknowledge support by the Vetenskapsr\aa det (Swedish Research
Council) through contract No. 638-2013-8993 and the Oskar Klein Centre for  Cosmoparticle Physics. M.L. acknowledges support from ASI through ASI/INAF Agreement 2014-024-R.1 for the Planck LFI Activity of Phase E2. O.M. is supported by PROMETEO II/2014/050, by the Spanish
Grant FPA2014--57816-P of the MINECO, by the MINECO Grant
SEV-2014-0398 and by the European Union’s Horizon 2020
research and innovation programme under the Marie Skłodowska-Curie grant
agreements 690575 and 674896. O.M. would like to thank the Fermilab Theoretical Physics Department for its hospitality. E.G. is supported by NSF grant AST1412966.
 S.H. acknowledges support by NASA-EUCLID11-0004, NSF AST1517593 and NSF AST1412966.
\end{acknowledgments}

\end{document}